\documentclass[preprint]{aastex63}

\graphicspath{{./}{figures/}}

\usepackage{color, soul}
\usepackage{graphicx}
\usepackage{subfigure}
\usepackage{rotating}
\usepackage{amsmath}

\begin{document}

\title{X-ray Constraints on the Hot Gas Content of Early-type Galaxies in Virgo}

\author{Meicun Hou}
\affiliation{School of Astronomy and Space Science, Nanjing University, Nanjing 210046, China}
\affiliation{Key Laboratory of Modern Astronomy and Astrophysics (Nanjing University), Ministry of Education, Nanjing 210046, China}
\email{houmc@smail.nju.edu.cn}

\author{Zhiyuan Li}
\affiliation{School of Astronomy and Space Science, Nanjing University, Nanjing 210046, China}
\affiliation{Key Laboratory of Modern Astronomy and Astrophysics (Nanjing University), Ministry of Education, Nanjing 210046, China}
\email{lizy@nju.edu.cn}

\author{Christine Jones}
\affiliation{Smithsonian Astrophysical Observatory, Cambridge, MA 02138, USA}

\author{William Forman}
\affiliation{Smithsonian Astrophysical Observatory, Cambridge, MA 02138, USA}

\author{Yuanyuan Su}
\affiliation{Department of Physics \& Astronomy, University of Kentucky, Lexington, KY 40506, USA}

\begin{abstract}
We present a systematic study of the diffuse hot gas around early-type galaxies (ETGs) residing in the Virgo cluster, 
based on archival {\it Chandra} observations. 
Our representative sample consists of 79 galaxies with low-to-intermediate stellar masses ($M_* \approx 10^{9-11}\rm~M_\odot$), a mass range that has not been extensively explored with X-ray observations thus far.
We detect diffuse X-ray emission in only eight galaxies and find that in five cases a substantial fraction of the detected emission can be unambiguously attributed to truly diffuse hot gas, based on their spatial distribution and spectral properties.
For the individually non-detected galaxies, we constrain their average X-ray emission by performing a stacking analysis, finding a specific X-ray luminosity of $L_{\rm X}/M_* \sim 10^{28}{\rm~erg~s^{-1}~M_{\odot}^{-1}}$, which is consistent with unresolved stellar populations. 
The apparent paucity of truly diffuse hot gas in these low- and intermediate-mass ETGs may be the result of efficient ram pressure stripping by the hot intra-cluster medium. 
However, we also find no significant diffuse hot gas in a comparison sample of 57 field ETGs of similar stellar masses, for which archival {\it Chandra} observations with similar sensitivity are available. 
This points to the alternative possibility that galactic winds evacuate
the hot gas from the inner region of low- and intermediate-mass ETGs, regardless of the galactic environment. 
Nevertheless, we do find strong morphological evidence for on-going ram pressure stripping in two galaxies (NGC\,4417 and NGC\,4459).
A better understanding of the roles of ram pressure stripping and galactic winds in regulating the hot gas content of ETGs, 
invites sensitive X-ray observations for a large galaxy sample.

\end{abstract}


\section{Introduction} \label{sec:intro}

It has been established that X-ray emission from normal galaxies have two primary components, namely, discrete stellar sources and diffuse hot gas with temperatures of $\sim 10^{6-7}$~K \citep[e.g.,][]{Trinchieri1985,Sarazin2001,Revnivtsev2008}. For massive early-type galaxies (ETGs), the X-ray emission is dominated by hot gas atmospheres accumulated from mass loss of old stellar populations (e.g., stellar winds from evolved stars, planetary nebulae, and Type Ia supernovae [SNe]), accretion of the intergalactic medium (IGM), as well as from mergers of small galaxies.  
In ETGs of low-to-intermediate masses, the X-ray emission is typically dominated by stellar sources, consisting of low-mass X-ray binaries (LMXBs), and to a less extent, cataclysmic variables (CVs) and coronally active binaries (ABs) \citep[e.g.,][]{Revnivtsev2006}. 
Hot gas may also exist in the shallow gravitational potential of these systems, likely in the form of a galactic-scale outflow driven by Type Ia SNe and/or an active galactic nucleus (AGN) \citep[e.g.,][]{David2006,Li2011,Bogdan2012a}.
The hot gas is a crucial component of the galactic ecosystem, mediating the transport of energy, mass, and metals between the interstellar medium (ISM) and the IGM.

Given the expectation that the hot gas content has a dependency on the host galaxy properties, the correlation between diffuse X-ray luminosity ($L_{\rm X}$) and stellar mass ($M_*$; also often represented by the integrated stellar luminosity, e.g., $L_{\rm K}$) has been extensively studied \citep[e.g.,][]{OSullivan2001,Boroson2011,Su2015,Goulding2016}. 
These studies generally found that the diffuse X-ray Luminosity (i.e., resolved X-ray sources subtracted) spans over two orders of magnitude for a given host stellar mass, especially in massive ETGs with $M_{\star} \gtrsim 10^{11}\rm~M_{\sun}$.
Several interpretations have been proposed for this large dispersion in $L_{\rm X}$: dependency on galaxy structures \citep[e.g.,][]{Mathews2006,Sarzi2013,Su2015}, energy feedback from AGN activity and/or supernova (SNe) heating \citep{David2006}, and environmental effects
\citep[e.g.,][]{White1991,Brown2000,Sun2007,Jeltema2008,Mulchaey2010,Wagner2018}.
In particular, in a dense environment such as galaxy groups and clusters, ram pressure stripping can effectively remove  hot gas coronae around individual galaxies, at a rate that depend on the galaxy's position and relative velocity \citep[e.g.,][]{Roediger2015a,Vijayaraghavan2015}.

Thus far, most  studies of the X-ray scaling relations have focused on massive ETGs ($M_{\star} \gtrsim 10^{11}\rm~M_{\sun}$), while the hot gas content in low- to intermediate-mass ETGs ($M_{\star} \lesssim 10^{11}\rm~M_{\sun}$) has received relatively little attention.
This is partly owing to a selection effect, in the sense that in small galaxies the X-ray emission from discrete sources generally overwhelms that of the hot gas, rendering a robust detection of the latter difficult. 
Previous studies of low- to intermediate-mass ETGs are either based on small-size samples \citep[e.g.,][]{Bogdan2012a} or case studies of particular interest (e.g., NGC\,4435, \citealp{Machacek2004}, NGC\,5102, \citealp{Kraft2005}). 
A systematic study of the hot gas content in a sizable sample of low-to-intermediate mass ETGs is still absent.

The Virgo cluster, the nearest galaxy cluster and the host of $\sim$1600 known member galaxies with more than half being ETGs \citep{Kim2014}, is a uniquely promising laboratory to study the environmental effects on the hot gas content in galaxies spanning a wide range of mass and size.
Indeed, a disturbed X-ray morphology has been seen in a number of Virgo spirals \citep[e.g.,][]{Tschoke2001,Machacek2004,Wezgowiec2012,Ehlert2013} and ellipticals \citep[e.g.,][]{Biller2004,Randall2004,Machacek2006,Randall2008,Kraft2011,Kraft2017,Paggi2017,Wood2017,Su2019}, providing strong
evidence for on-going ram-pressure stripping in these galaxies. 
Nevertheless, we still lack precise knowledge about the hot gas content in the majority of normal galaxies in Virgo. 

We are thus motivated to carry out a systematic study of diffuse X-ray emission in a sample of low- to intermediate-mass ETGs in Virgo, based on archival {\it Chandra} observations.
{\it Chandra}'s superb angular resolution and sensitivity for point source detection facilitate the removal of bright LMXBs in the vicinity of individual ETGs (\citealp{Hou2017}, hereafter Paper I), which is crucial to minimize stellar contamination in the faint unresolved X-ray emission and to reveal the truly diffuse hot gas.
We show below that the available {\it Chandra} observations allow us to probe the diffuse X-ray emission in galaxies as small as $M_* \sim 10^{9}\rm~M_{\sun}$.
To evaluate the anticipated environmental effects on the hot gas content, we also extract a sample of field ETGs with comparable stellar masses from the {\it Chandra} archive, which provide a statistically meaningful comparison with the Virgo ETGs. 

The rest of the paper is organized as follows. Section~\ref{sec:data} describes our sample selection and data reduction. The analysis of diffuse X-ray emission in both the Virgo and field ETGs are presented in Section~\ref{sec:analysis}. The implications of our results are discussed in Section~\ref{sec:discussion}, followed by a summary in Section~\ref{sec:summary}. 
Throughout this work, we adopt a uniform distance of 16.5 Mpc (1$\arcsec$ corresponds to 80 pc; \citealp{Mei2007}) for all sources and galaxies in Virgo. The line-of-sight depth of Virgo introduces an uncertainty of $\lesssim$ 10\% in the derived luminosities, which should have little effect on our results. Errors are quoted at a 1$\sigma$ confidence level, unless otherwise stated.




\section{Data Preparation} \label{sec:data}
\subsection{Sample Selection and Galaxy Properties}\label{subsec:sample}
Similar to Paper I, we started from a parent sample defined in a {\it Chandra Large Program}, AGN Multiwavelength Survey of Early-Type Galaxies in the Virgo Cluster \citep[AMUSE--Virgo,][]{Gallo2008,Gallo2010}, which consists of 100 ETGs spanning three orders of magnitude in stellar mass. 
In paper I, we analyzed 80/100 galaxies each with $\sim$5 ks snapshot exposures with the Advanced CCD Imaging Spectrometer (ACIS). 
The same exposure of 5 ks ensured a quasi-uniform limiting luminosity (0.5--8 keV) of $\sim2\times 10^{38}{\rm~erg~s^{-1}}$ for source detection, which was essential for studying the statistical properties (spatial distribution and luminosity function) of the X-ray point sources around the individual ETGs (Paper I).
Here, we slightly modify the sample with the aim of studying the diffuse X-ray emission.
First, we excluded the 10 most massive galaxies with a stellar mass $> 10^{11} {\rm~M_{\odot}}$, which have been extensively studied in the literature.
We further discarded a few galaxies in the close vicinity of M87, M49 or M86, because these galaxies are severely contaminated by the extended hot gas halo of the nearby giant elliptical and thus have an exceptionally high local background. This removed 6 galaxies close to M87 ($<35'$), 2 galaxies close to M49 ($<10'$) and 1 galaxy close to M86 ($<12')$.
We also excluded two galaxies observed in ACIS sub-array mode, the small field-of-view (FoV) of which is not suitable for studying the diffuse emission.

\begin{figure*}\centering
\includegraphics[scale=0.46, angle=0]{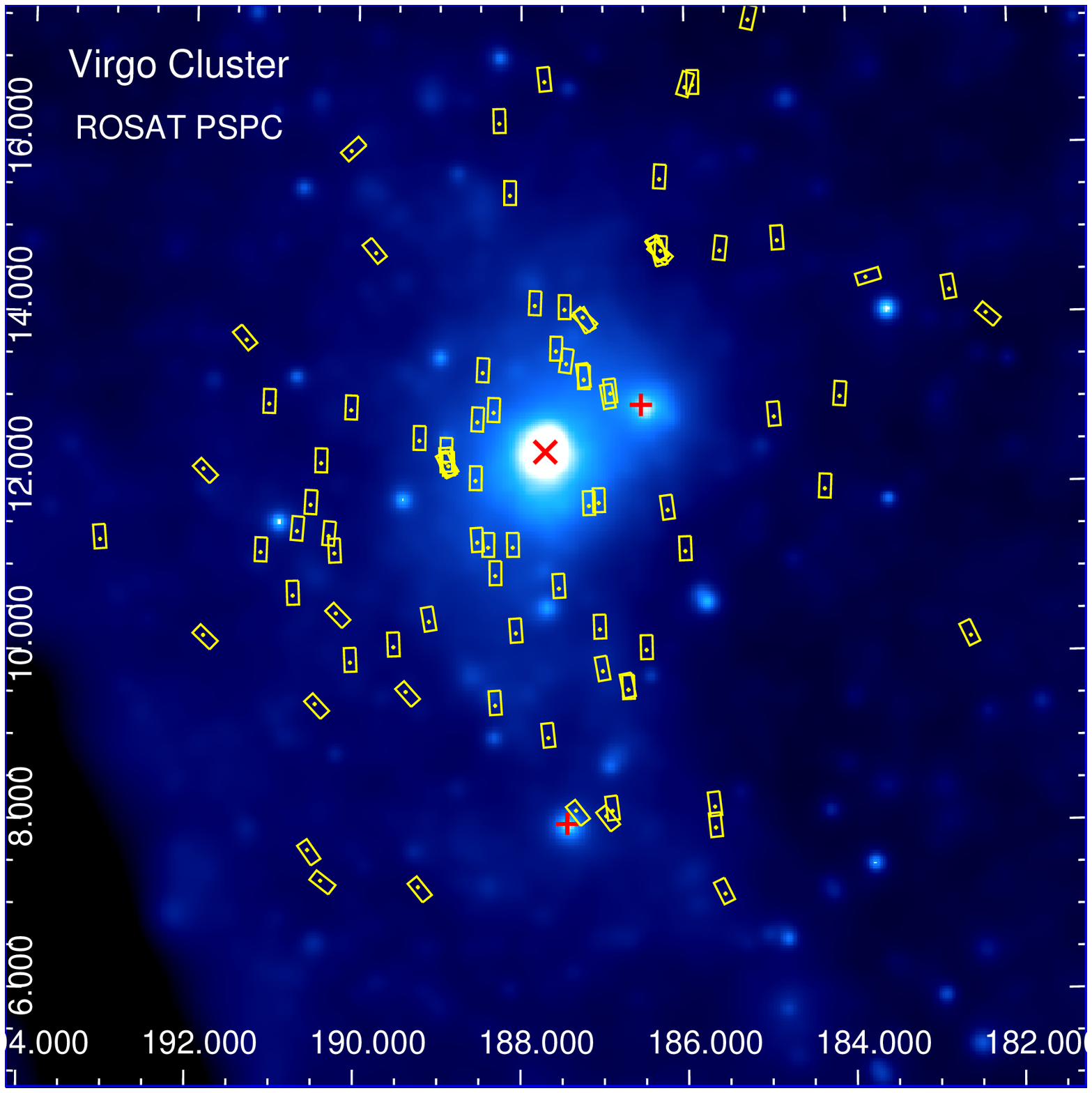}
\includegraphics[scale=0.5, angle=90]{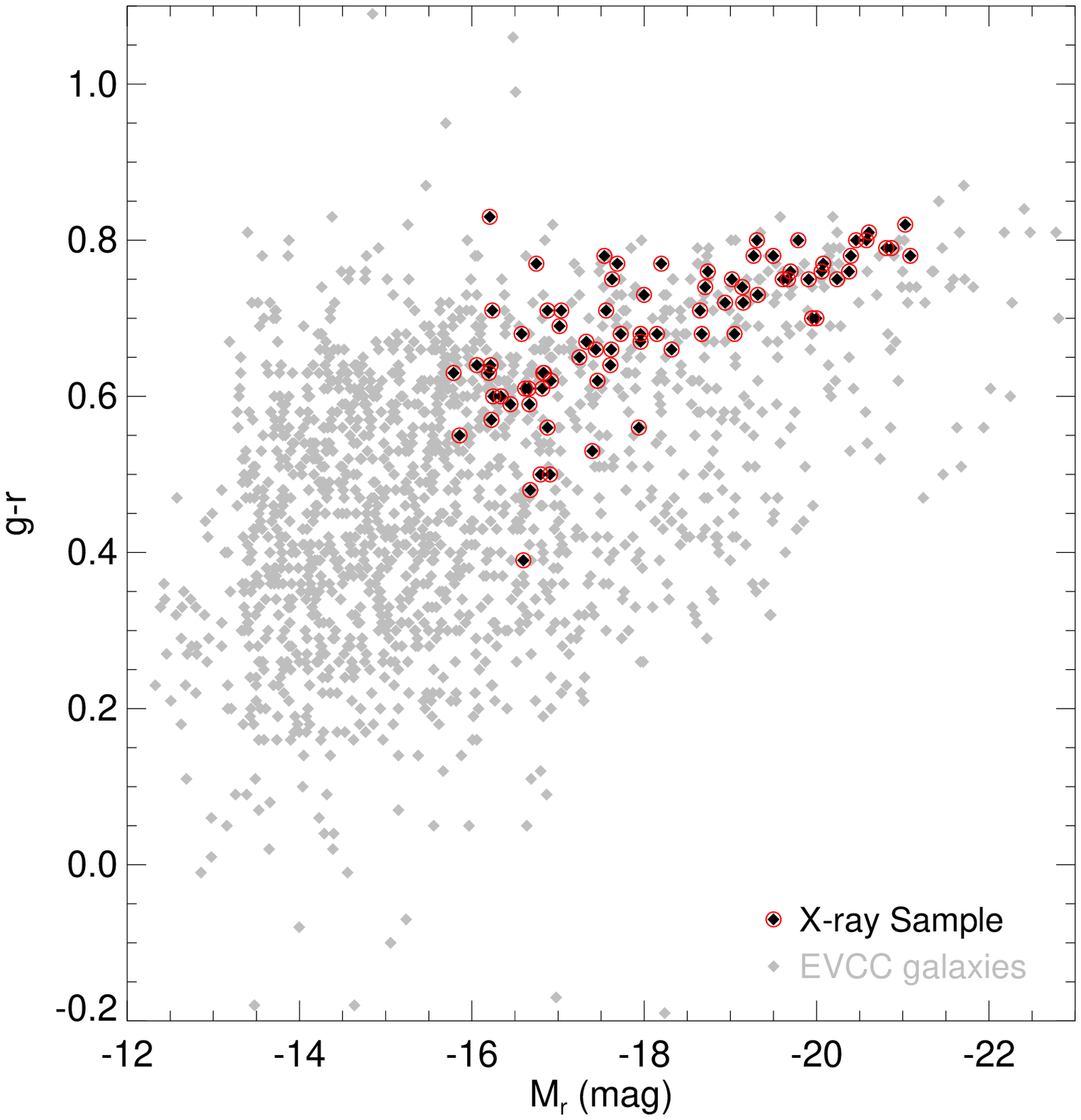}
\caption{Our sample of Virgo early-type galaxies. {\it Left}: Footprint of the 79 Virgo ETGs on the ROSAT all-sky survey image of the Virgo cluster \citep{Bohringer1994}. The individual {\it Chandra}/ACIS fields are outlined by the yellow boxes, with the center of each galaxy marked by a dot. The center of M87 is marked by a red cross, while the centers of M49 and M86 are marked by a red `+'. 
{\it Right}: $g - r$ color versus $r$-band absolute magnitude of 
$\sim$1600 Virgo member galaxies (grey diamonds) compiled in the Extended Virgo Cluster Catalog \citep{Kim2014}.
The 79 ETGs studied in this work are highlighted by red circles.
}
\label{fig:sample}
\end{figure*}

Our final sample thus consists of 79 low- to intermediate-mass ETGs in Virgo.
Among them, ten galaxies have more than one ACIS observations, for which we incorporate all the available data to maximize the signal-to-noise ratio (S/N) for the unresolved emission. 
The total on-target exposure of these galaxies ranges between 9 -- 132 ks. One galaxy, VCC\,1231, has a single exposure of $\sim$ 30 ks, and the remaining 68 galaxies have a single exposure of $\sim$ 5 ks.
A large set of galaxies with a quasi-uniform exposure is essential for a stacking analysis (see Section \ref{subsec:stack}).
The sky positions of these 79 ETGs, illustrated in the left panel of Figure~\ref{fig:sample}, nicely sample the Virgo cluster. 
The right panel of Figure~\ref{fig:sample} plots the $g-r$ color-magnitude diagram of all ETGs compiled in the Extended Virgo Cluster Catalog \citep[EVCC,][]{Kim2014}, highlighting the 79 ETGs studied here.
Almost all of our targets lie along the red sequence of quiescent galaxies.
Their basic information, including galaxy name, celestial coordinates and observation ID, are presented in Table~\ref{tab:info}.

To reveal the potential environmental effects on the diffuse X-ray emission, 
we employed a {\it comparison sample} of 57 field ETGs, which were originally defined by a {\it Chandra} large program, AMUSE-Field \citep{Miller2012} and were also used in Paper I.
Since the distances of these field ETGs span from 10.1 Mpc to 27.3 Mpc, 
the individual ACIS exposures were designed to vary in such a way that they reach a quasi-uniform limiting luminosity of $\sim2\times 10^{38}{\rm~erg~s^{-1}}$ for point sources, i.e., identical to that of the Virgo ETGs. 
It is noteworthy that none of these field ETGs has additional {\it Chandra} observations than what AMUSE-Field program provides.

The effective radius ($R_e$) of each Virgo ETG was adopted from the {\it Hubble Space Telescope} ({\it HST})/Advance Camera for Surveys Virgo Cluster Survey \citep[ACSVCS;][]{Ferrarese2006}, in which $R_e$ was derived from a fit to the $g$-band surface brightness profile using a S\'{e}rsic model. 
For one galaxy, VCC\,1030, a reliable fit was not possible due to the presence of a small dusty disk. Thus, we estimated its $R_e$ based on the mean scaling relation calibrated among the ACSVCS galaxies \citep[Eq.~21 in][]{Ferrarese2006}, ${\rm log}R_{e} = (-0.055 \pm 0.023)(M_{\rm B}+18) + (1.14 \pm 0.03)$, where $M_{\rm B}$ is the $B$-band absolute magnitude of a given galaxy and $R_e$ is in units of arcsec.
For VCC\,1512, a dIrr/dE transition object, and VCC\,575, which has a stellar disk or ring in its center, the surface brightness profile and related integrated quantities were deemed unreliable. Thus, we also estimated a new $R_e$ value for each of them based on the above empirical relation.
The $R_e$ values of the Virgo ETGs range from $3\farcs7$ to $58\farcs1$ (0.3--4.6 kpc), with a median value of $12\farcs8$ ($\sim 1.0$ kpc).
Due to the lack of {\it HST} imaging for most AMUSE-Field galaxies, we also applied the empirical scaling relations from the ACSVCS survey to determine a S{\'e}rsic profile, in particular the S{\'e}rsic index and effective radius, for each field galaxy, based on the $B$-band magnitudes tabulated in \citep{Miller2012}.
The resultant $R_e$ values of the field galaxies range from 0.6 to 1.4 kpc, with a median value of 0.8 kpc.
We have also estimated $R_e$ values for the field ETGs using another empirical relation based on the $B$-band absolute magnitude \citep[][Eq.~16 therein]{Graham2008} 
and found consistent results.
The stellar masses of the Virgo ETGs and field ETGs were taken from \citet{Gallo2010} and \citet{Miller2012}, respectively. 

\subsection{X-ray Data Reduction}\label{subsec:X-ray data}
For all the selected ACIS fields, we downloaded and reprocessed the archival data using CIAO v4.8 and the calibration files CALDB v4.7.0, following the standard procedure\footnote{http://cxc.harvard.edu/ciao/}. 
All relevant observations had the aimpoint placed on the S3 CCD, and we only included data from the S3 and S2 CCDs.
We examined the light curve of each observation and found that the instrumental background was contaminated by significant particle flares in only a few cases, for which we filtered the affected time intervals. The cleaned exposure of each galaxy is listed in Table~\ref{tab:info}.
For each observation, we produced counts and exposure maps on the natal pixel scale ($0 \farcs 492 {\rm~pixel^{-1}}$) in the 0.5--2, 2--8 and 0.5--8 keV bands. 
The exposure maps were weighted by a fiducial incident spectrum, which is an absorbed power-law with a photon-index of 1.7 and an absorption column density $N_{\rm H}$ = $10^{21} {\rm~cm^{-2}}$, which is typical for X-ray point sources.
We also generated the corresponding instrumental background maps from the ``stowed background'' data, after normalizing with the 10--12 keV count rate.
For galaxies with more than one observations, we calibrated their relative astrometry by matching the centroid of commonly detected point sources, using the CIAO tool {\it reproject\_aspect}. The count and exposure maps of individual observations were then reprojected to a common tangential point, i.e., the optical center of the galaxy, to produce combined images of enhanced S/N.

Following the procedures detailed in Paper I, we performed source detection across the FoV of each galaxy in each of the three energy bands. 
This results in a total of $\sim$1400 point-like sources detected in and around the 79 Virgo ETGs,   
which are composed of AGNs and active galaxies in the cosmic background, AGNs and LMXBs associated with the individual ETGs, as well as intra-cluster X-ray sources (Paper I).
Similarly, we detected $\sim$1030 sources in and around the 57 field ETGs. 
To study the diffuse X-ray emission, we removed pixels falling within two times the 90\% enclosed-energy radius (EER) around each detected point source.
These pixels are consistently removed from the counts, exposure and instrumental background maps, and no artificial counts are introduced to fill these pixels in the following quantitative analysis.

\section{Analysis and results} \label{sec:analysis}
\subsection{Detection of Diffuse X-ray Emission} \label{subsec:diffuse}

\begin{figure*}\centering
\includegraphics[scale=0.52, angle=90]{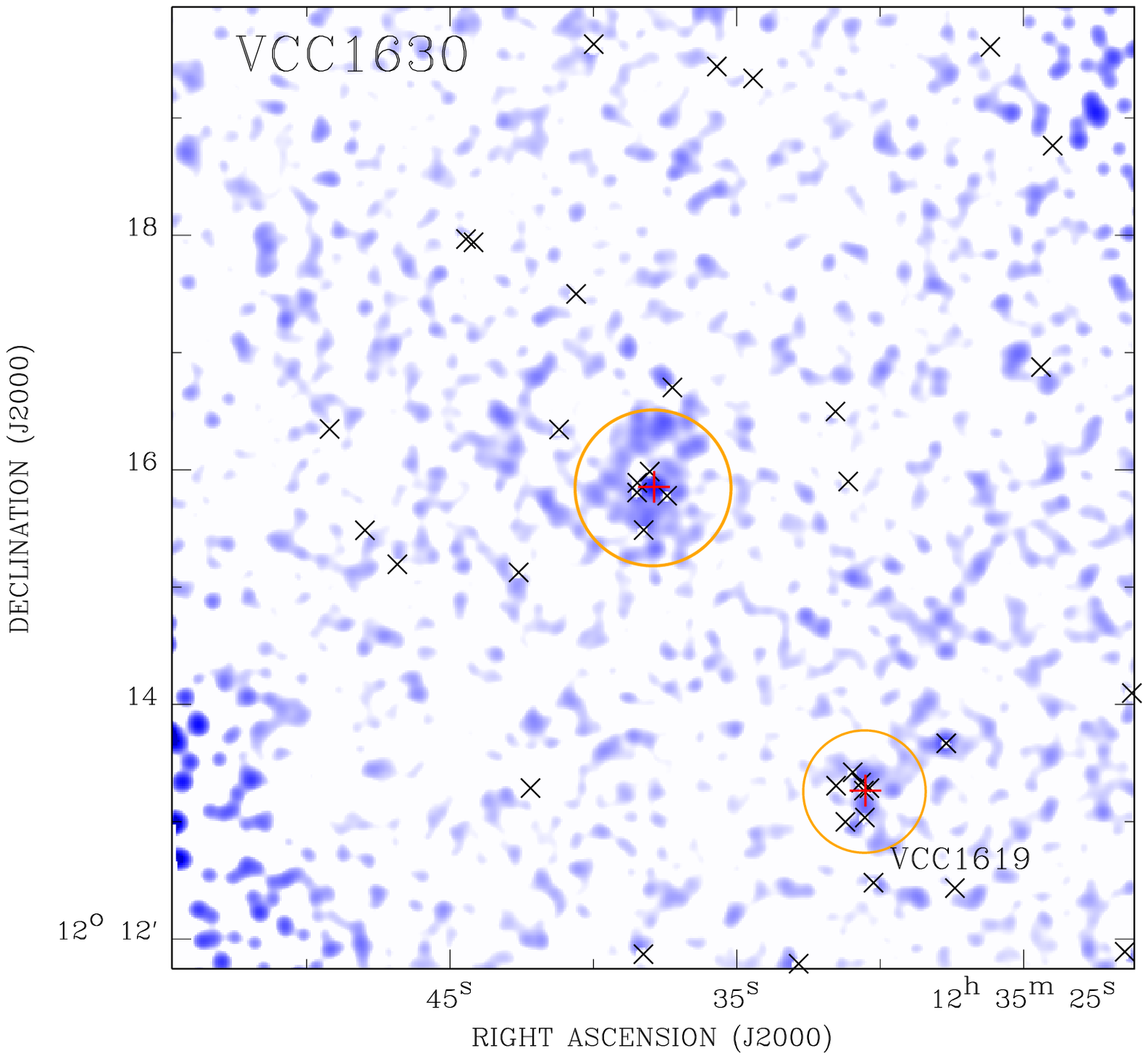}
\includegraphics[scale=0.46, angle=90]{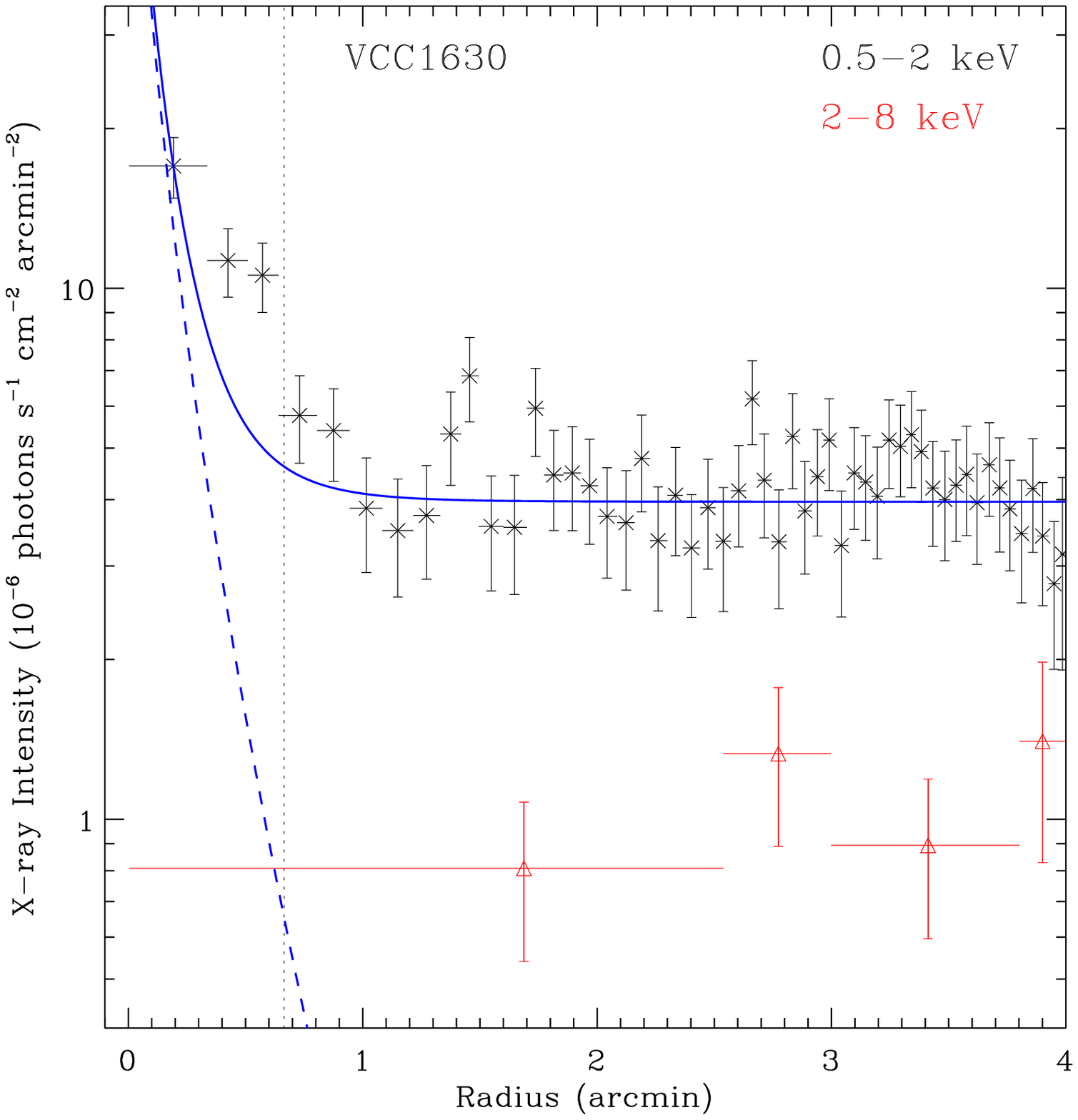}
\includegraphics[scale=0.52, angle=90]{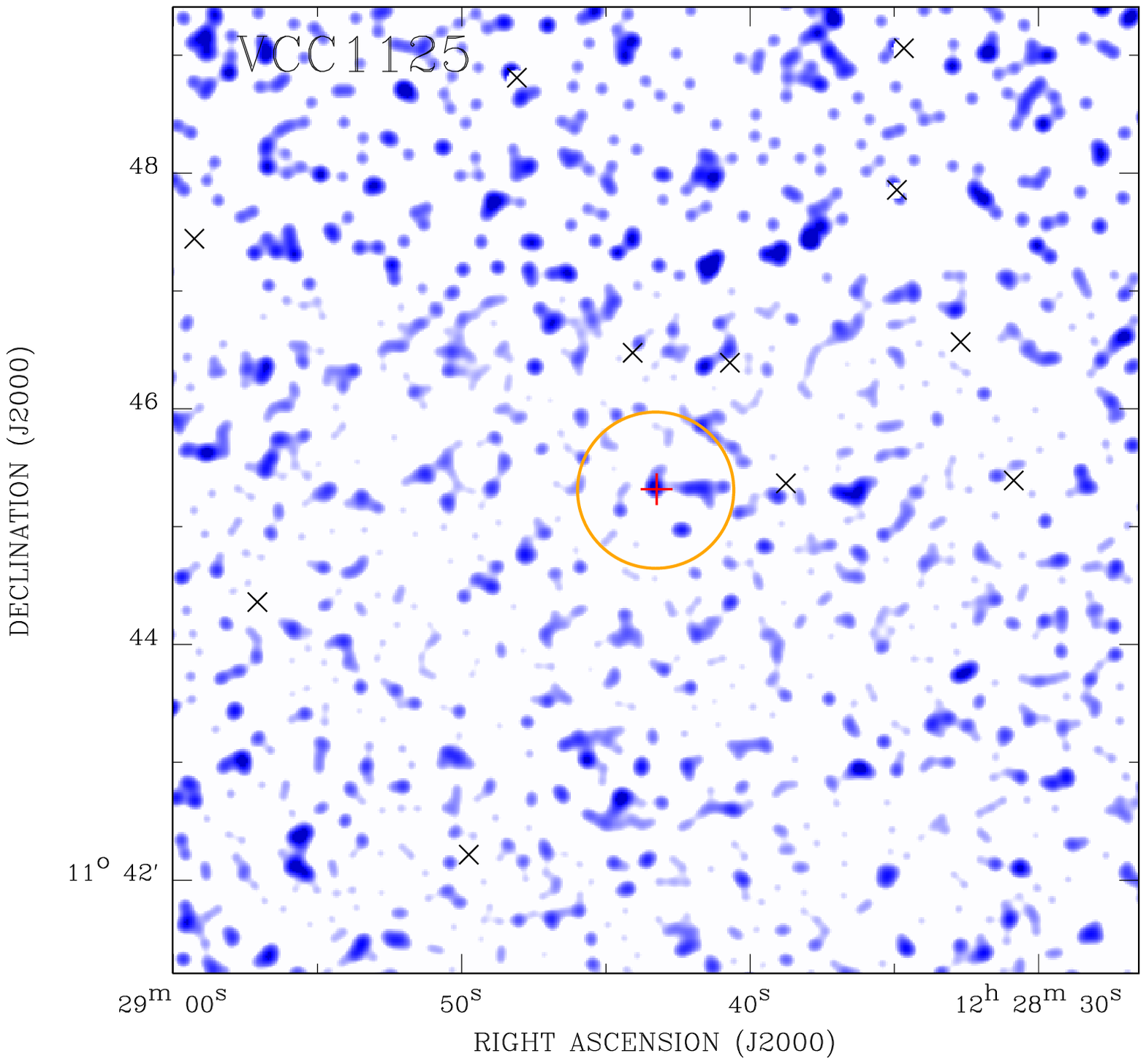}
\includegraphics[scale=0.46, angle=90]{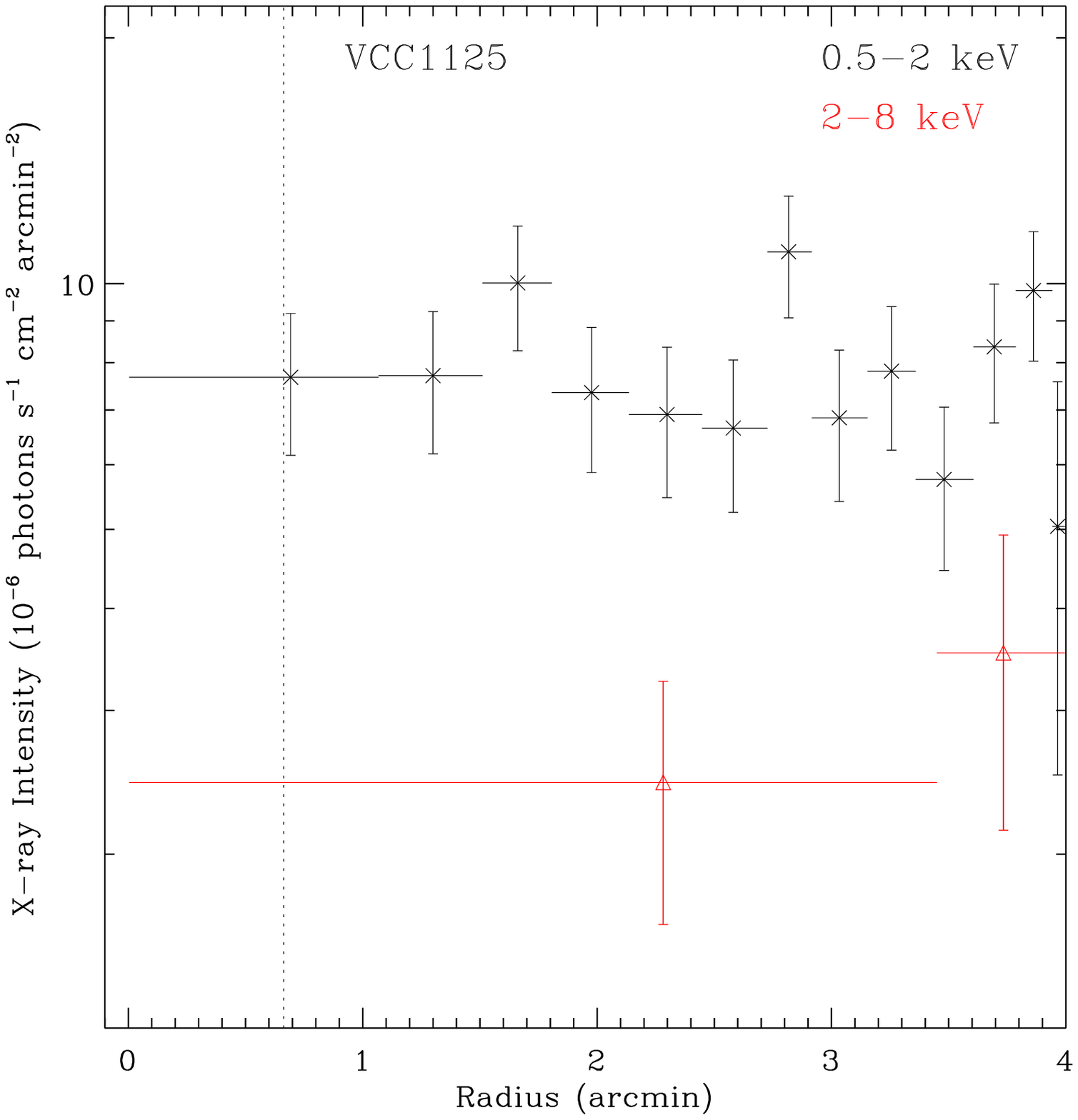}
\caption{{\it Upper left}: 0.5--2 keV flux map of VCC\,1630, a representative early-type galaxy with extend X-ray emission. The map is smoothed with a Gaussian kernel of 15 ACIS pixels ($\sim 7\farcs5$). The red `+' sign marks the center of the galaxy, whereas the black crosses mark the positions of the removed point sources.
The orange circle has a radius of 3$R_e$. 
The ETG VCC\,1619 falling within the FoV is also marked.
{\it Upper right}: Instrumental background-subtracted, exposure-corrected radial intensity profiles of the unresolved X-ray emission in the 0.5--2 keV (black) and 2--8 keV (red) bands, adaptively binned to achieve a S/N greater than 3. The blue dashed line represents the $g$-band starlight S\'{e}rsic profile from the ACSVCS survey, normalized to match the innermost data point of the 0.5--2 keV profile. 
The blue solid line shows the sum of the S\'{e}rsic profile and a constant local background.
The vertical dotted line marks the position of 3$R_e$.
{\it Lower panels}: similar to the upper panels but for VCC\,1125, a galaxy representative of those Virgo ETGs showing no significant diffuse emission.
}
\label{fig:radial}
\end{figure*}

We first examine the flux maps of the 79 Virgo ETGs in the 0.5--2 keV band, which are produced by subtracting the corresponding instrumental background maps and divided by the exposures maps. Due to low surface brightness nature of most galaxies, these maps have been
smoothed by a Gaussian kernel of 15 ACIS pixels ($\sim 7\farcs5$) for a better visulization.  
From the flux maps, extended features are clearly visible around a few galaxies, while the rest, the majority of galaxies show no obvious diffuse emission. Figure~\ref{fig:radial} shows an example for each case: VCC 1630 exhibits extended emission out to at least three times its effective radius, whereas VCC\,1125, a galaxy of similar stellar mass, shows no significant residual emission after source subtraction. 
We further construct the source-subtracted radial intensity profile in the 0.5--2 keV and 2--8 keV bands for each galaxy, out to a projected radius of $4^\prime$ ($\sim$19.2 kpc) from the galactic center. 
This is the maximum radius at which the ACIS FoV provides full azimuthal coverage, and is sufficiently large compared to the typical angular size of our sample galaxies. 
The radial intensity profiles of VCC\,1630 and VCC\,1125 are again shown as examples in the right panels of Figure~\ref{fig:radial}. 
Evidently, VCC\,1630 shows a centrally enhanced 0.5--2 keV profile extending out to $\gtrsim3 R_e$, whereas the profile of VCC\,1125 is consistent with a flat, local background. 
For VCC\,1630, we further compare its 0.5--2 keV radial intensity profile with the best-fit $g$-band S{\'e}rsic profile. The latter, obtained from the ACSVCS survey \citep{Ferrarese2006}, represents the galaxy's starlight distribution. It can be seen that the 0.5--2 keV profile is significantly more extended than the starlight, suggesting the presence of truly diffuse hot gas. 
The little bump in the  X-ray intensity profile at $\sim 3\farcm2$ is likely due to another Virgo ETG VCC\,1619 located southwest of VCC\,1630.
Due to the limited S/N, both galaxies show no significant signals in the 2--8 keV band above the local background.
This is generally true for the relatively shallow observations used in this work, hence we shall not further discuss the 2--8 keV band. 

\begin{figure*}\centering
\includegraphics[scale=0.4, angle=0]{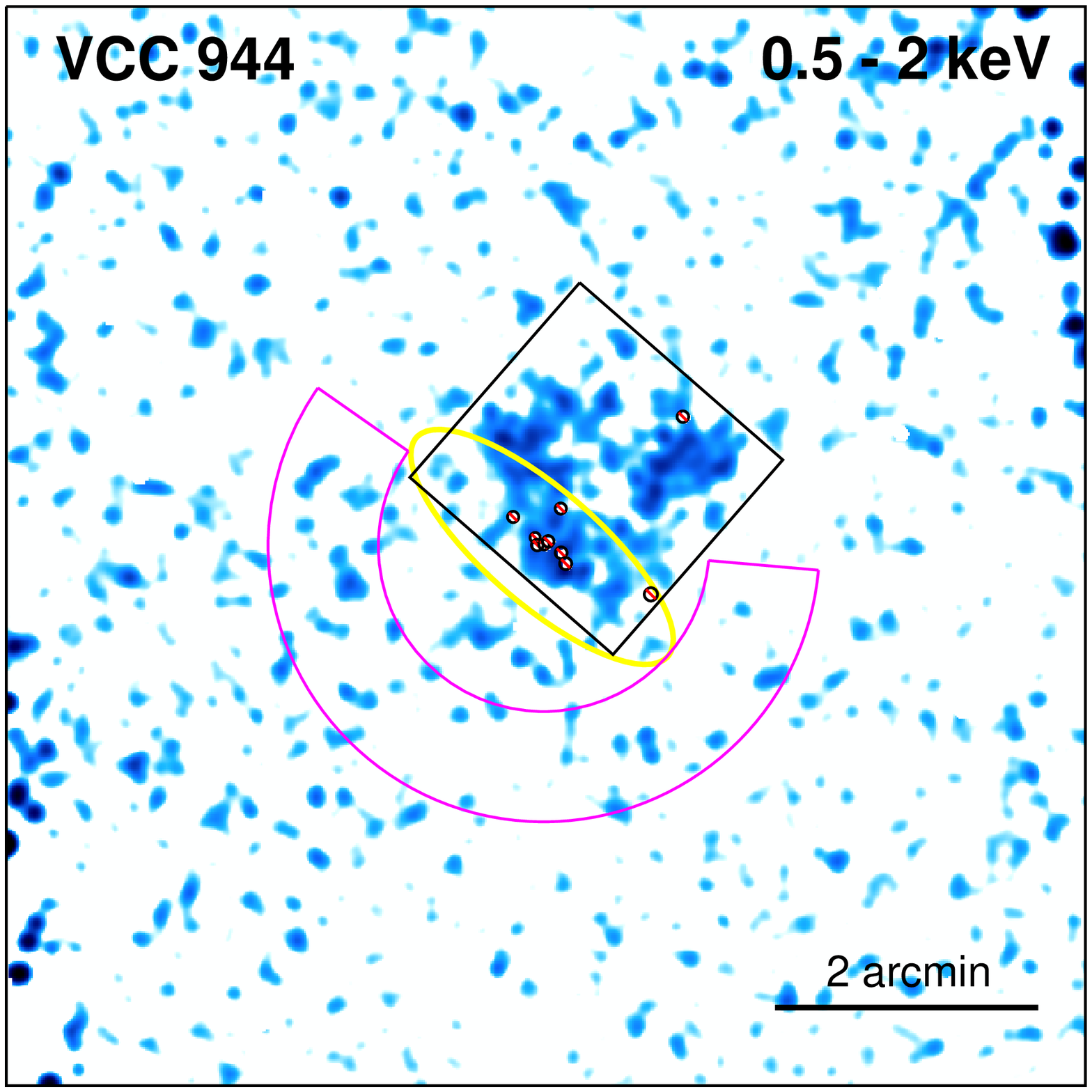}
\includegraphics[scale=0.4, angle=0]{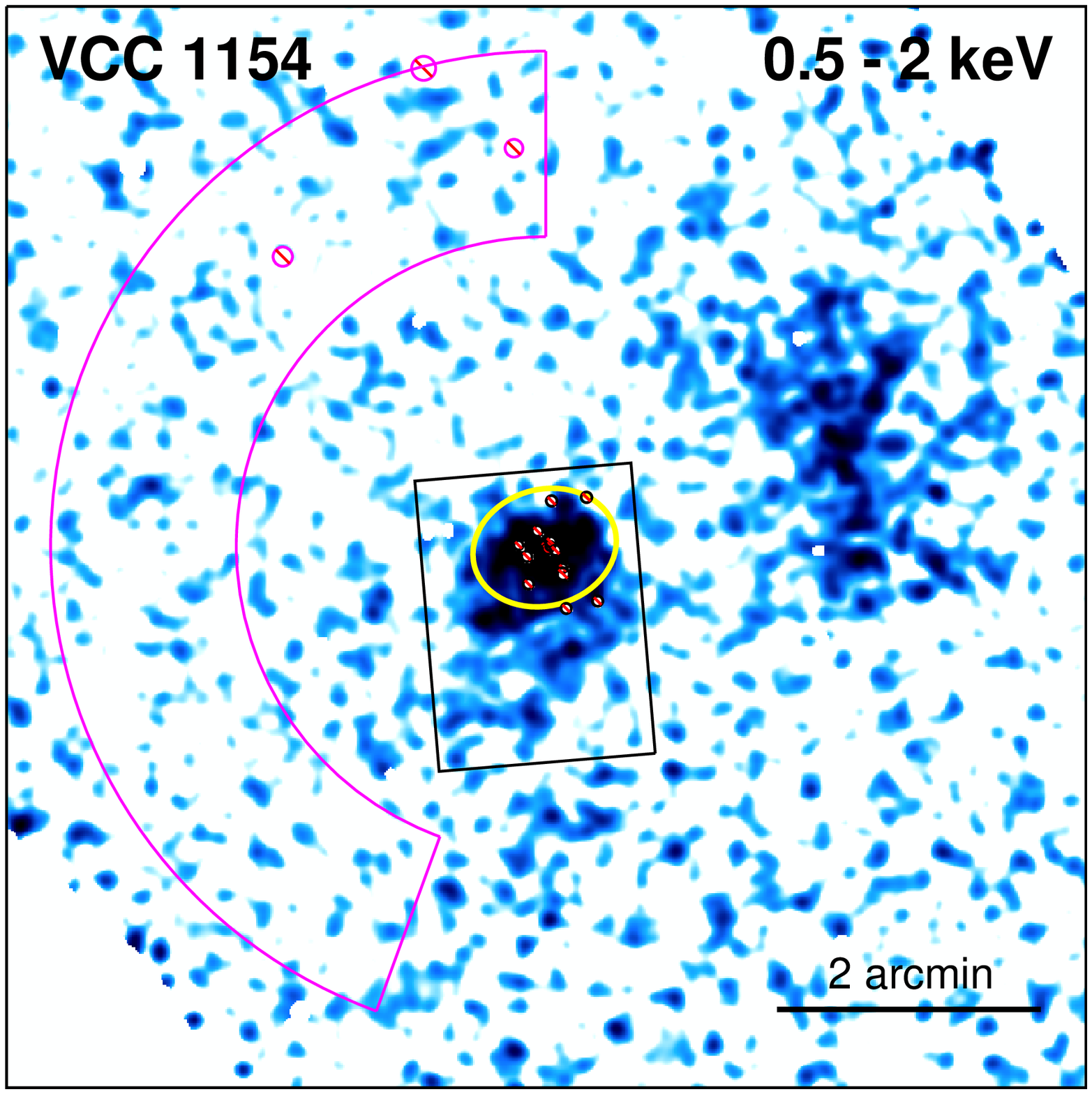}
\caption{The 0.5-2 keV flux maps of VCC\,944 and VCC\,1154. The maps are smoothed with a Gaussian kernel of 15 ACIS pixels ($\sim 7\farcs5$). Diffuse X-ray features are present beyond 3$R_e$ in both galaxies. The yellow ellipse outlines the optical extent of the host galaxy. The source extraction region is marked by the black box, while the local background region is marked by the magenta sector. In both regions the detected point sources are masked.
}
\label{fig:fov_detected}
\end{figure*}

We uniformly quantify the S/N of the 0.5--2 keV diffuse emission for each galaxy, as follows.
The region of diffuse emission is defined as a circle with a radius of 3$R_e$, masking the detected point sources within two times of the 90\% EER.
The corresponding local background region is defined as an annulus with an inner radius of 5$R_e$ and an outer radius of 8$R_e$. 
For the few largest galaxies, this outer radius slightly exceeds $4^\prime$, in which case the area of the incomplete annulus is taken into account.
While this choice of source and background regions works well for the majority of galaxies, two cases demand extra care, which are illustrated in Figure~\ref{fig:fov_detected}.
VCC\,944, with ${\rm log}M_* = 10.4$, exhibits significant diffuse emission extending well beyond 3$R_e$ to the north/northwest. 
A similar situation is found in VCC\,1154 (${\rm log}M_* = 10.4$)  with significant diffuse emission extending beyond 3$R_e$ to the south. Besides, another prominent feature of diffuse emission is seen well beyond 3$R_e$ to the northwest. This feature appears detached from the diffuse emission within 3$R_e$.
We have examined the SDSS and HST images of these two galaxies but found no obvious optical counterparts to the extended X-ray features. The nature of these two cases will be further discussed in Section~\ref{sec:discussion}.
Thus we adjust the source and background regions of VCC 944 and VCC 1154 to accommodate their irregular morphology, as illustrated in Figure~\ref{fig:fov_detected}.

Next, for a given galaxy, we employ the CIAO tool {\it aprate}, which applies the Bayesian approach for Poisson statistics
at the low-count regime, to calculate the net photon flux and bounds.
In this calculation the local background, effective exposure and enclosed energy fraction are taken into account.
A galaxy is regarded as a firm detection of diffuse emission, if the 3$\sigma$ lower limit of its net photon flux given by {\it aprate} is greater than zero.
In such a case the 1$\sigma$ errors of the net photon flux and the S/N of the signal are reported (Table~\ref{tab:info}). For those galaxies regarded as non-detections, we report the 3$\sigma$ upper limit of the net photon flux. 

A total of 8 galaxies are found to have S/N $> 3$.
Indeed all these 8 galaxies (VCC\,759, VCC\,778, VCC\,944, VCC\,1030, VCC\,1062, VCC\,1154, VCC\,1231, VCC\,1630) clearly show diffuse signals in their 0.5--2 keV flux maps.
Notably, they all have a stellar mass ${\rm log}M_* > 10.0$. 
A further examination of the radial intensity profiles of these galaxies finds that in five cases (VCC\,944, VCC\,1030, VCC\,1154, VCC\,1231, VCC\,1630) the 0.5--2 keV emission is more extended than the starlight, suggesting the presence of truly diffuse hot gas. 
The morphology of VCC\,944, VCC\,1154, and VCC\,1630 are shown in Figure \ref{fig:fov_detected} and Figure \ref{fig:radial}. In VCC\,1030, the extended X-ray emission is lopsided to the northeast, suggesting possible interactions with the nearby massive spiral galaxy NGC\,4438 (\citealp{Machacek2004}). In VCC\,1231, the diffuse X-ray emission shows a more symmetric morphology out to $\sim 3R_{e}$. 
Among these eight galaxies, five have multiple observations, which certainly help provide an enhanced S/N. 
Diffuse emission remains undetected in the other five galaxies with multiple observations, which notably all have ${\rm log}M_* < 10.0$.  

We note that additional Virgo member galaxies may fall within the ACIS FoV of our target galaxies (Paper I). For completeness, we also search for significant diffuse X-ray emission from such galaxies, provided that they are classified as an ETG in the EVCC \citep{Kim2014} and are 
brighter than 16 mag in the $r$-band, which is roughly the limiting magnitude of the AMUSE-Virgo galaxies. 
We thus identify 3 fields with two additional EVCC ETGs and 21 fields with one additional EVCC ETG. 
It turns out that the majority of these additional ETGs are located far from the target ETGs and show no obvious diffuse X-ray emission. Thus their presence will not affect the measurement of diffuse emission for the target ETGs. 
Only three fields deserve further attention. 
NGC\,4309 is located at $\sim 1\farcm5$ south of the target ETG VCC\,538,  
but neither galaxy has significant diffuse emission. 
NGC\,4461 is located at $\sim 3\farcm7$ southeast of the target ETG VCC\,1146. Extended emission is clearly seen from the 0.5--2 keV flux map and there is a bright nuclear source in NGC\,4461. However, our spectral analysis (Section~\ref{subsec:spectra}) indicates that this apparently  diffuse emission is totally dominated by the unresolved stellar population and requires no truly diffuse hot gas. 
Lastly, a lenticular galaxy NGC\,4477 is located at $\sim 5\farcm2$ southeast of the target ETG VCC\,1283 and is close to the edge of the FoV. Diffuse X-ray emission is clearly seen from the flux map of this galaxy, which has been studied in detail by \citet{Liyj2018}, who found a pair of symmetric cavities around the galactic center, possibly due to AGN activity.

\subsection{Stacking Non-detections} \label{subsec:stack}

\begin{figure*}\centering
\includegraphics[width=0.9\textwidth,, angle=0]{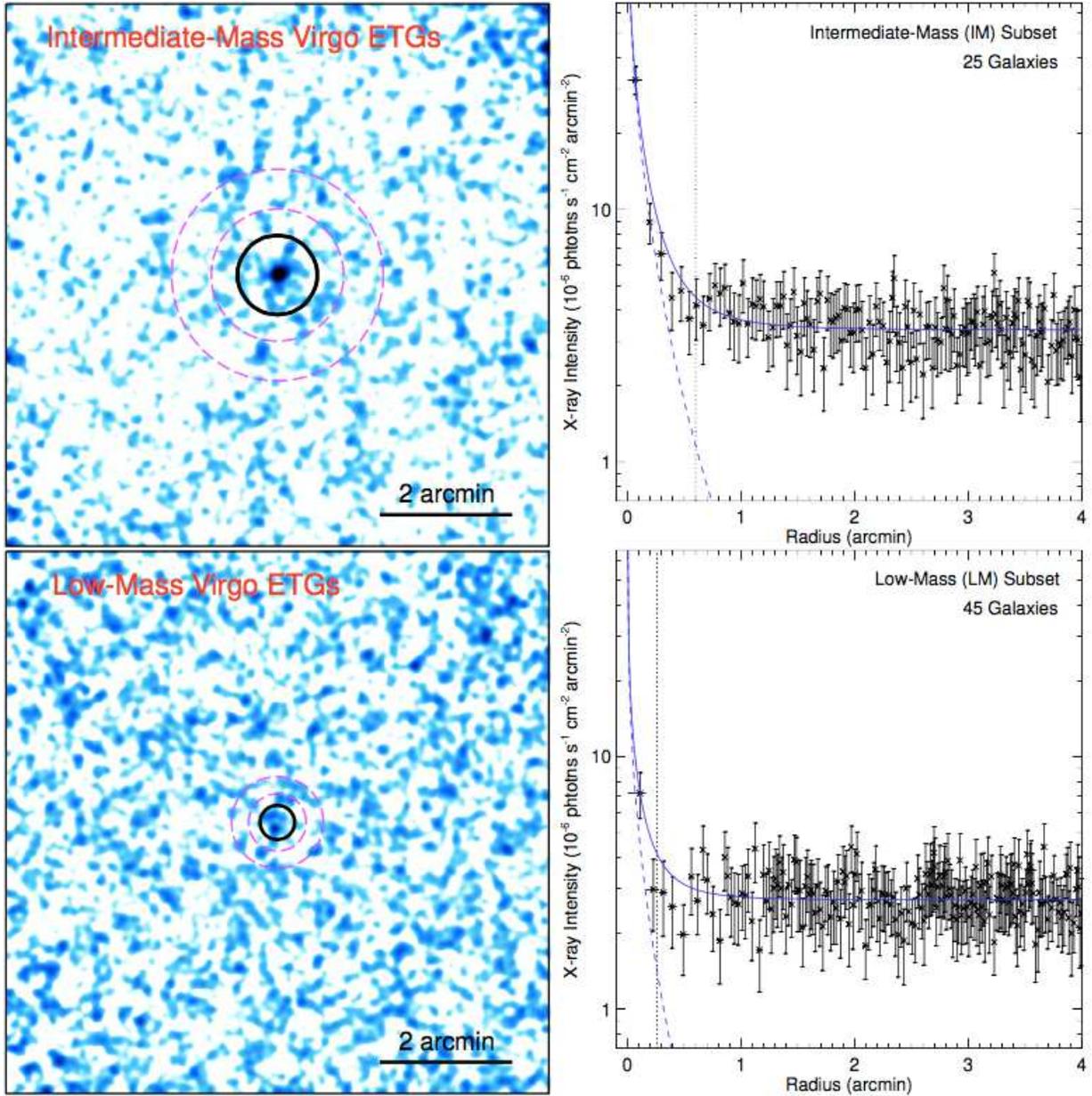}
\caption{The stacked 0.5--2 keV flux map ({\it left panels}) and averaged radial intensity profile ({\it right panels}) of individually non-detected Virgo ETGs, divided into the intermediate-mass bin ({\it upper panels}) and low-mass bin ({\it lower panels}). 
The flux maps are smoothed with a Gaussian kernel of 15 ACIS pixels ($\sim 7\farcs5$), and the intensity profiles are adaptively binned to achieve a S/N greater than 3.
The black circle marks the signal extraction region, 
within which the S/N of the diffuse emission is evaluated.
The magenta dashed annulus represents the local background region.
In the right panels, the black dotted vertical line marks the radius with which the signal is extracted.
The blue dashed line represents the averaged S\'{e}rsic profile, normalized to match the innermost data point. 
The blue solid line shows the sum of the S\'{e}rsic profile and a constant local background.
}
\label{fig:radial_stack}
\end{figure*}

Based on the X-ray photometric results (Section~\ref{subsec:diffuse}), the majority of our sample ETGs, especially the low-mass ones, do not have detectable diffuse X-ray emission. 
Thus, we perform a stacking analysis for the non-detected galaxies. 
Among the 71 non-detected Virgo ETGs, one galaxy, VCC\,2019 (= IC\,3735), shows two notable diffuse X-ray features at around 3$R_e$ to the south and southwest. A visual examination of the SDSS image indicates possible optical counterparts of these X-ray features, which have a photo-$z$ $\sim$ 0.3--0.4, thus they are more likely in the background (e.g., arising from a background galaxy cluster) rather than physically associated with VCC\,2019. Hence we remove this galaxy from the following stacking analysis. 
Moreover, if a galaxy had more than one observation, we only include the one with $\lesssim$ 5 ks exposure to ensure that the stacked signal is not dominated by any single galaxy.
The 27 additional member ETGs falling within the ACIS FoV (Section \ref{subsec:diffuse}) have a projected location far away from the target ETGs and have no significant diffuse X-ray emission, thus they will not affect the stacking results.

To probe the potential dependence of diffuse X-ray emission on the stellar mass, we divide the remaining 70 non-detected ETGs into two subsets according to their stellar masses: an intermediate-mass (IM) subset containing 25 galaxies that have $9.9 \leq {\rm log}M_* \leq 10.9$, and a low-mass (LM) subset containing 45 galaxies with $8.7 \leq {\rm log}M_* < 9.9$. 
The cut between the two mass bins is somewhat arbitrary, but we have verified that a slightly different mass cut does not affect our conclusion below.
We then produce stacked counts maps for these two mass bins, in which each galaxy is reprojected to the center of the map, again with the detected point sources masked. 
The exposure maps and instrumental background maps are stacked in a similar way. 
The 0.5--2 keV flux map of the two mass bins are presented in Figure~\ref{fig:radial_stack} (left panels).
Significant signals are evident near the center of these maps, especially in the IM bin. 
This is confirmed by the stacked radial intensity profile of each mass bin, shown in the right panels of Figure~\ref{fig:radial_stack}.
We also compare the X-ray intensity profile with an average starlight distribution, which is obtained by summing up the best-fit S{\'e}rsic profiles of individual galaxies in a given mass bin. 
No significant excess is seen over this average starlight distribution, indicating that the stacked 0.5--2 keV signals in both mass bins are dominated by the stellar contribution. 

Similar to what is done for the individual galaxies, we quantify the S/N of the stacked signal by extracting 0.5--2 keV counts from 
a circular region (marked in Figure~\ref{fig:radial_stack}), the radius of which (0\farcm6 for IM bin and 0\farcm26 for LM bin) is guided by the radial intensity profile to enclose the bulk signal. We have verified that the resultant net flux is insensitive to the exact choice of this radius. The local background is calculated from an annulus (also marked in Figure~\ref{fig:radial_stack}) with an area $\sim$ 4 times of the source region.
{\it Aprate} reports S/N $\approx 5.7$ for the IM bin and S/N $\approx 3.2$ for the LM bin.
Results of the stacking analysis are summarized in Table~\ref{tab:subset}.

\subsection{Diffuse X-ray Emission in the Field ETGs} \label{subsec:field}
For the comparison sample of field ETGs, we employ the same procedure to study their diffuse X-ray emission.
First, we produce the 0.5--2 keV flux map for each galaxy and construct the radial intensity profile.
Next, the S/N of each field galaxy is evaluated using {\it aprate}, for which we adopt the same criteria to define the source and background regions.

Based on the flux maps, radial intensity profiles, and S/N from {\it aprate}, 
we find that 7 (NGC\,1426, NGC\,1439, NGC\,3193, NGC\,3928, NGC\,4648, NGC\,5582, NGC\,5638) of the 57 field ETGs have a firm detection of diffuse X-ray emission, while the remaining 50 are non-detections (i.e., with S/N $<3$).
Similar to the case of the Virgo ETGs, those 7 field ETGs with detectable diffuse emission all have a stellar mass ${\rm log}M_* \gtrsim 10.0$. 
However, only one galaxy, NGC\,3193, has a more extended X-ray intensity profile than the starlight.
In another case, NGC\,3928, the diffuse emission is compact and concentrated within $\sim7\arcsec$ around a detected nuclear source, which might be the sign of an AGN-driven outflow, but the limited number of counts ($\sim25$) prevents us from drawing a firm conclusion. 
The other five detected field ETGs have a very moderate S/N, preventing a reliable assessment of their intrinsic morphology.


\begin{figure*}\centering
\includegraphics[width=0.95\textwidth, angle=0]{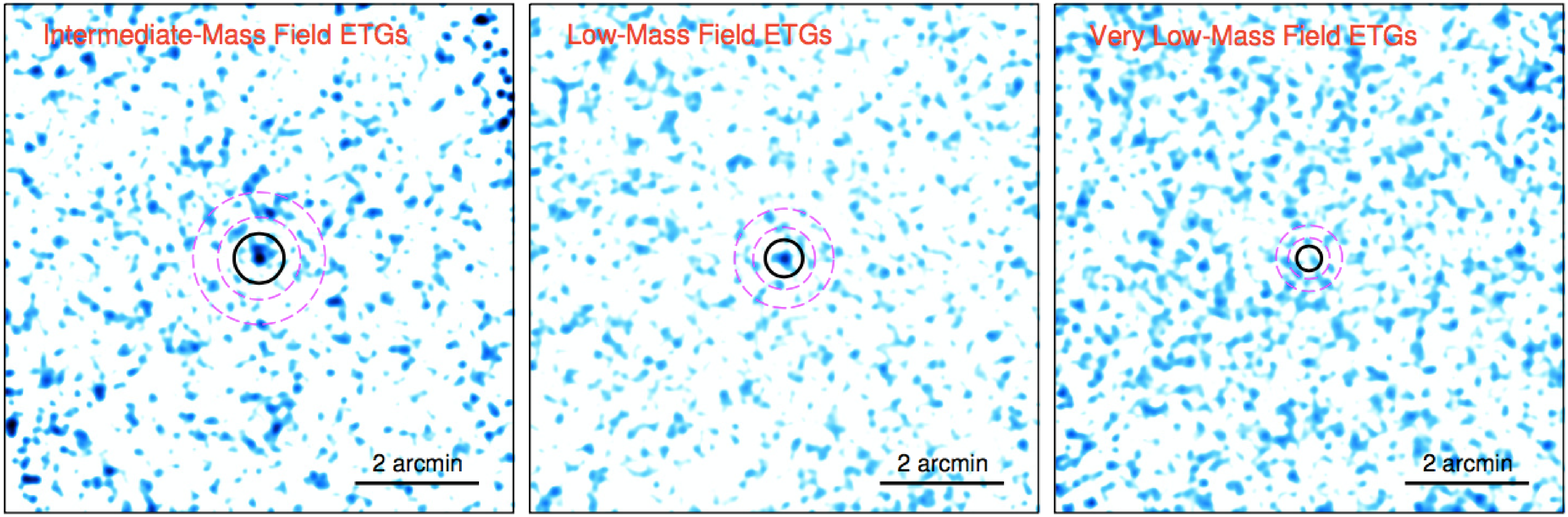}
\caption{The stacked 0.5--2 keV flux maps of individually non-detected {\it field} ETGs, divided in intermediate-mass, low-mass and very low-mass subsets, shown from left to right.
The black circle in each map marks the signal extraction region, 
within which the S/N of the diffuse emission is evaluated.
The magenta dashed annulus represents the local background region.
}
\label{fig:stack_field}
\end{figure*}

After visual examination of each case, we find that one galaxy, ESO\,540-014, is contaminated by a nearby bright point source. 
Therefore, we remove this galaxy from the subsequent stacking analysis.
We divide the remaining 49 field ETGs into three subsets according to their stellar masses: 
an intermediate-mass (IM) subset containing 7 galaxies that have $9.9 < {\rm log}M_* < 10.9$, a low-mass (LM) subset containing 15 galaxies with $8.7 \leq {\rm log}M_* < 9.9$, and a very low-mass (VLM) subset containing 27 galaxies with $7.7 \leq {\rm log}M_* < 8.7$. 
The first two subsets are meant to cover the same mass ranges as in the Virgo ETGs (Section~\ref{subsec:stack}). 
The VLM subset, on the other hand, has no counterpart in the Virgo ETGs. Nevertheless, it is valuable to examine the X-ray emission from these very low mass ETGs, which have received little attention in the literature.

We then produce the stacked 0.5--2 keV flux maps for these three mass bins, as shown in Figure~\ref{fig:stack_field}. 
In both the IM and LM bins,
signals are evident near the center of the flux map, but no significant emission can be seen in the map of the VLM bin.
Using a similar approach for the Virgo ETGs, we quantify the S/N for these three mass bins, finding a S/N of 3.6, 3.1 and 0.3 for the IM, LM and VLM bin (extracted from a circular region with radius of 0\farcm4, 0\farcm3 and 0\farcm2), respectively.

\subsection{Spectral Analysis} \label{subsec:spectra}

\begin{figure*}\centering
\includegraphics[scale=0.36, angle=270]{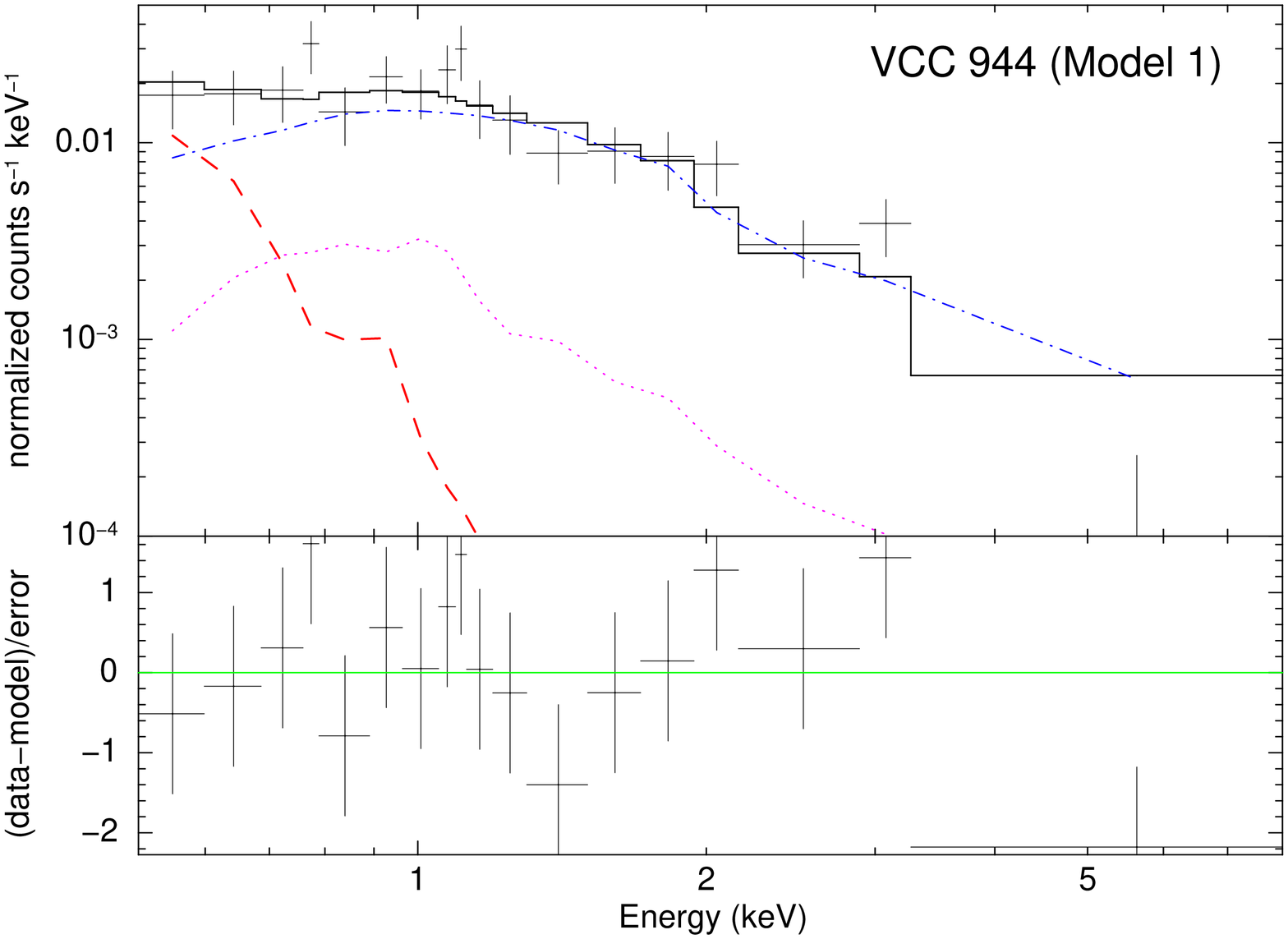}
\includegraphics[scale=0.36, angle=270]{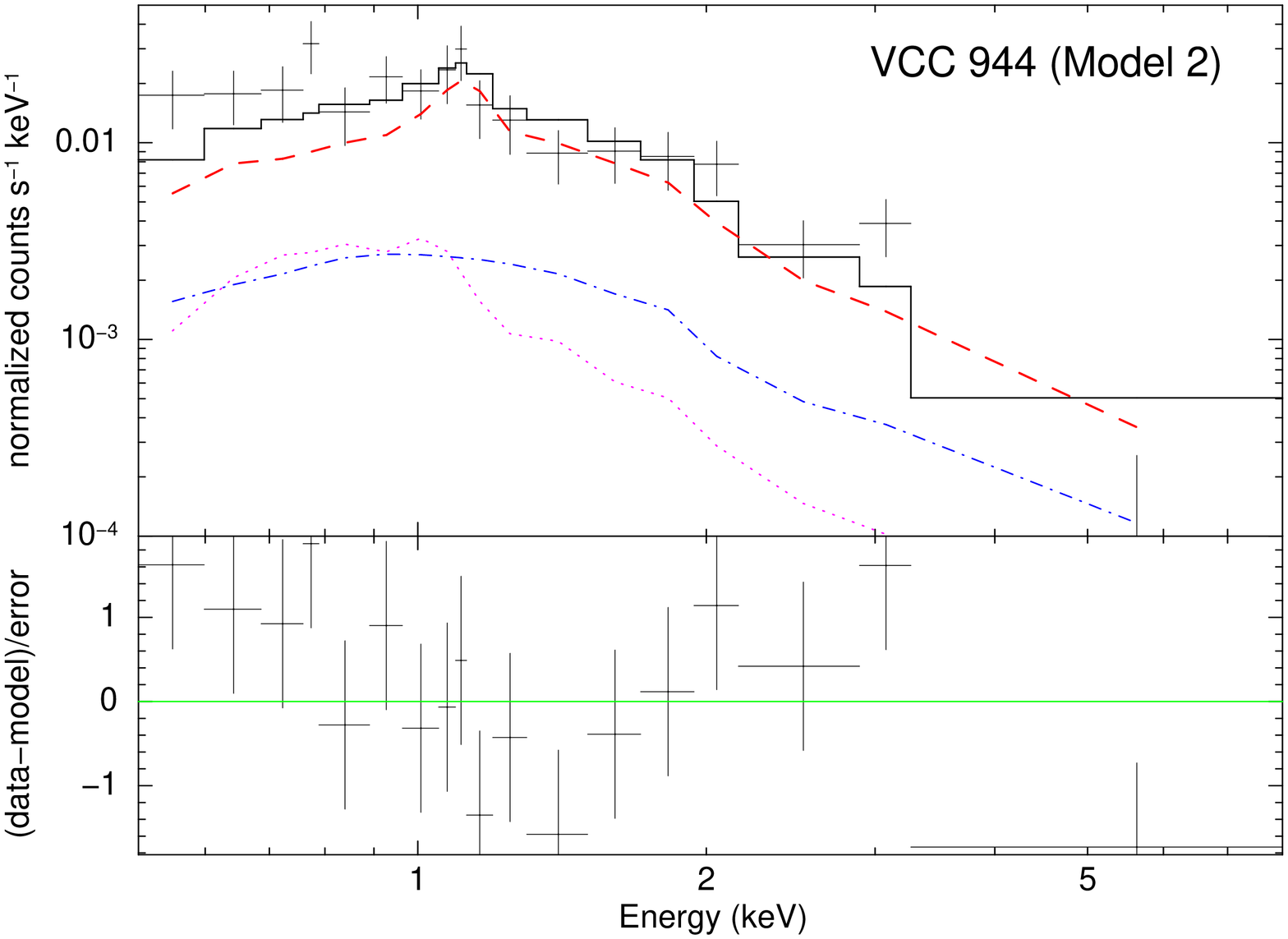}
\includegraphics[scale=0.36, angle=270]{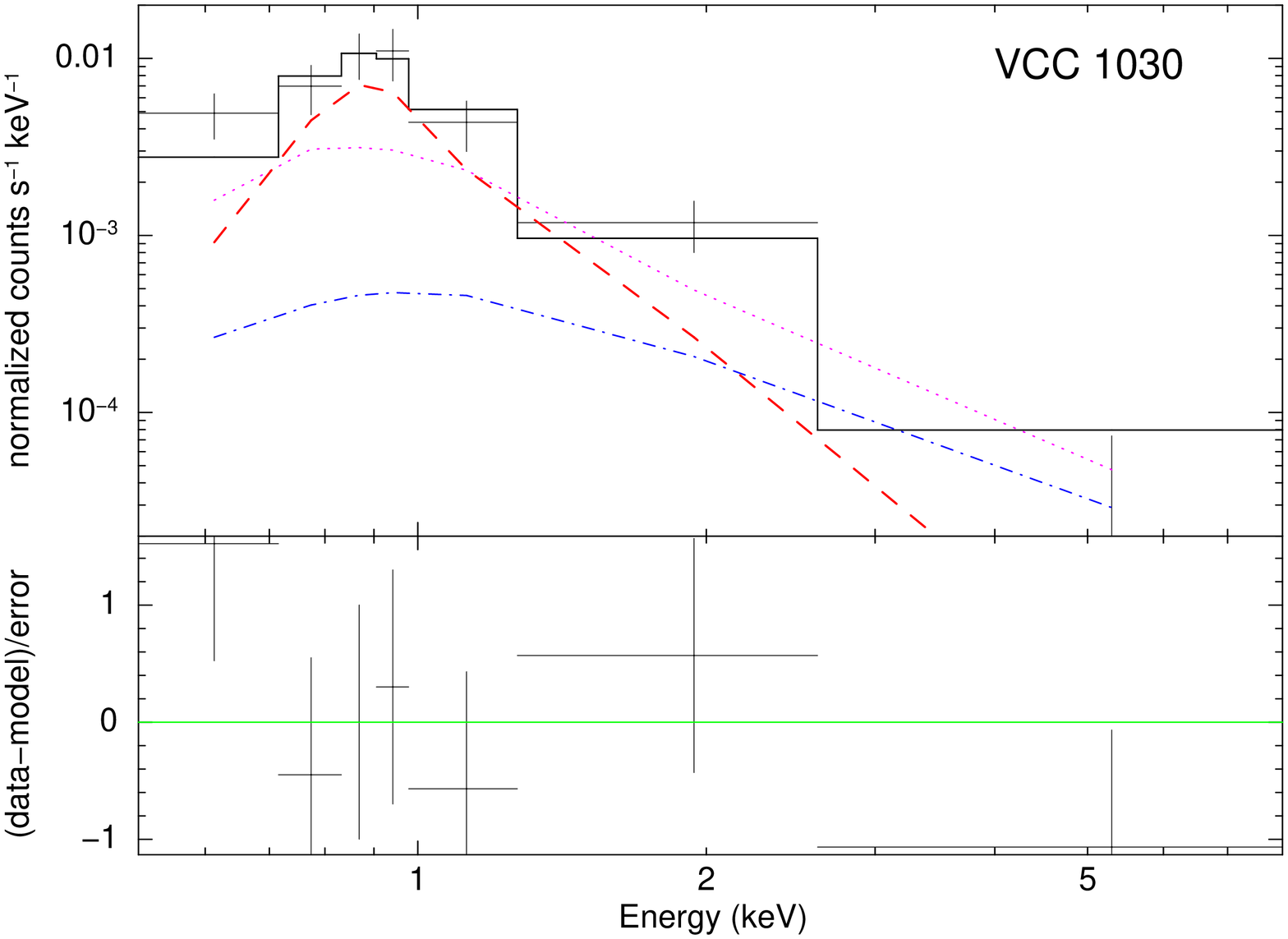}
\includegraphics[scale=0.36, angle=270]{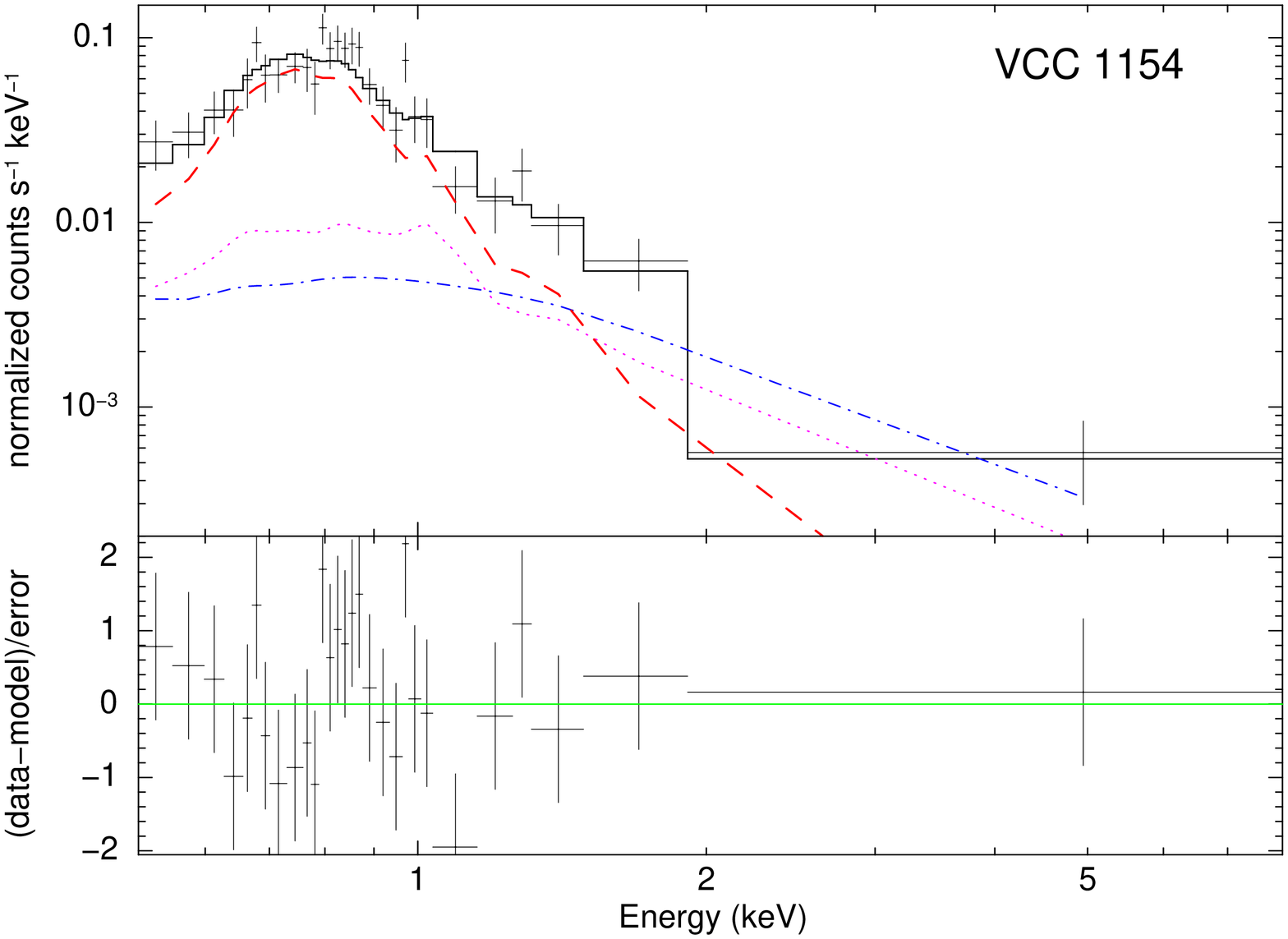}
\includegraphics[scale=0.36, angle=270]{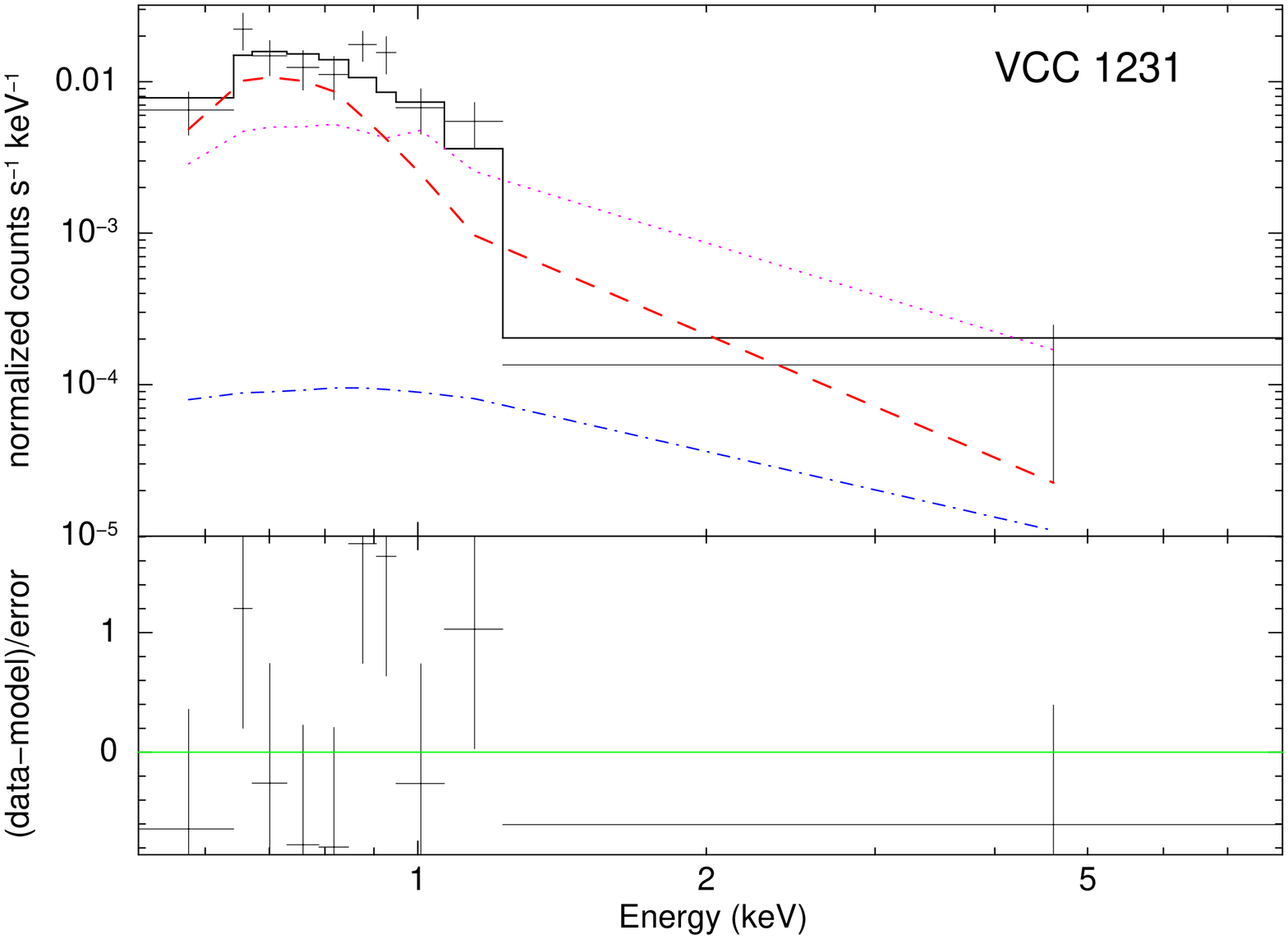}
\includegraphics[scale=0.36, angle=270]{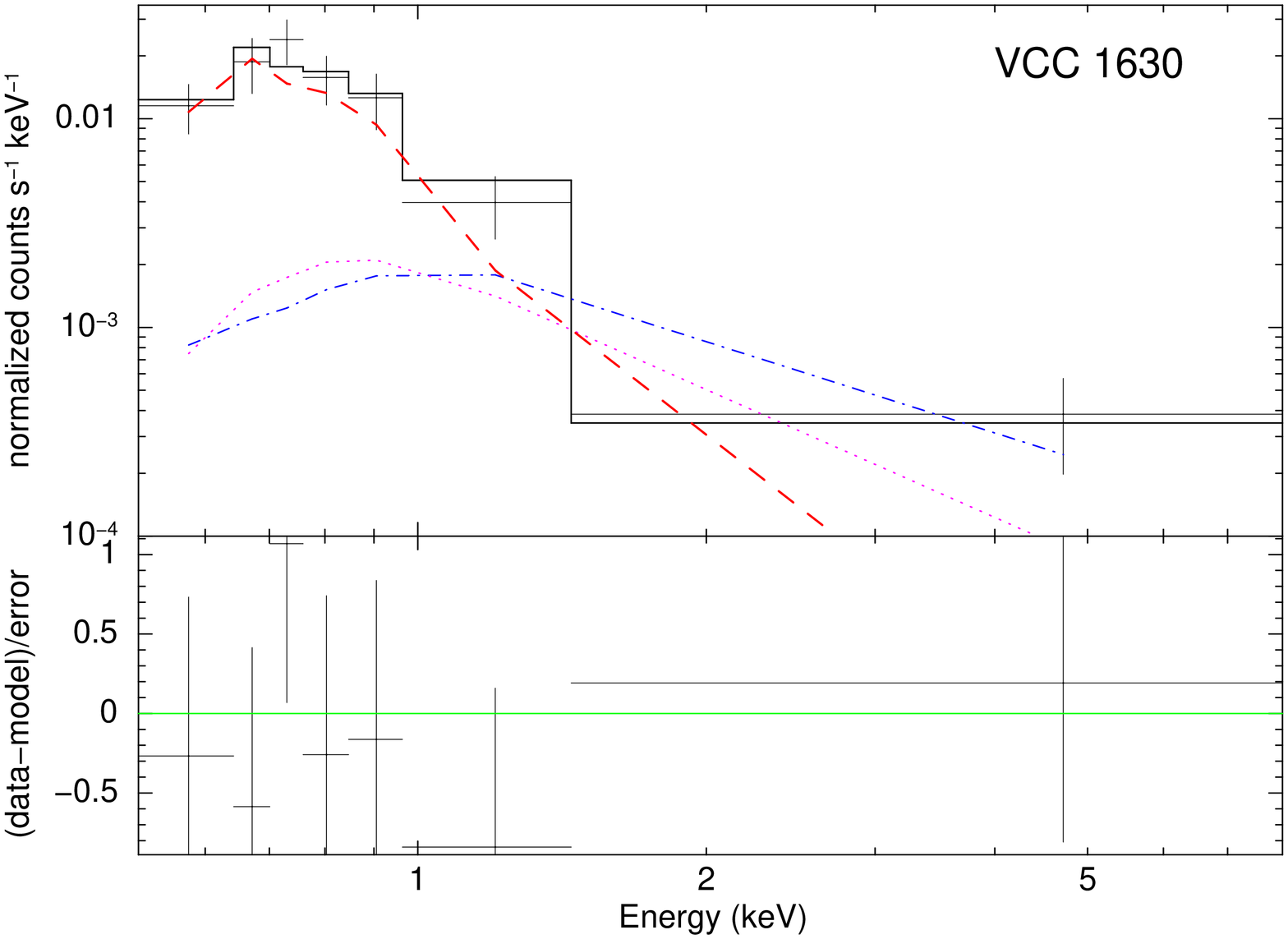}
\caption{Spectral modeling for the five galaxies with $>$100 net counts. 
The spectral data have been grouped to achieve a signal-to-noise ratio of 3. The fitted model is {\it $phabs\times(lnvmkl+powerlaw+apec)$}, where the {\it lnvmkl} (magenta dotted line) represents the contribution from CVs+ABs, the power-law (blue dot-dashed line) accounts for unresolved LMXBs, and the {\it apec} (red dashed line) accounts for truly diffuse hot gas. 
For each galaxy the lower panel plots the value of $\chi$.
For VCC\,944, VCC\,1154 and VCC\,1630, the spectra from two or multiple observations are jointly fitted, but we show here the combined spectrum for clarity. Two spectral models have been attempted for VCC\,944 (the first row): in the first model, the contribution from the unresolved LMXBs is set free, while in the second model this component is fixed at the expected level.
}
\label{fig:spectra}
\end{figure*}

Five (VCC\,944, VCC\,1030, VCC\,1154, VCC\,1231, VCC\,1630) of the eight Virgo ETGs with detected diffuse X-ray emission have sufficient net counts ($> 100$) for a meaningful spectral analysis. 
We note that all these five galaxies exhibit a 0.5--2 keV radial intensity profile more extended than the starlight (Section~\ref{subsec:diffuse}).
Using the CIAO tool {\it specextract}, we extract spectra from the same source region as defined in Section~\ref{subsec:diffuse} for each observation of these five galaxies, along with the Response Matrix File, Ancillary Response File and the corresponding background spectrum from the aforementioned background region. 

As discussed in Section~\ref{subsec:diffuse}, the apparently diffuse X-ray emission from these galaxies may contain two physical components, one from a collection of unresolved stellar populations, namely, LMXBs, CVs and ABs, and the other from truly diffuse hot gas.
In particular, CVs and ABs with typical X-ray luminosities of $10^{28}-10^{34}{\rm~erg~s^{-1}}$ cannot be individually resolved outside the Milky Way, but their total number and collective X-ray luminosity have been inferred to be linearly proportional to stellar mass in normal stellar environments \citep{Revnivtsev2007,Ge2015}. 

Hence, we apply a composite model to fit the spectra.
For CVs and ABs, we adopt a phenomenological model that provides a reasonable fit to the unresolved X-ray spectra of M32 and three dwarf elliptical galaxies around M31 \citep{Revnivtsev2007,Ge2015},
which are thought to be dominated by CVs and ABs.
Specifically, the CV+AB component assumes a log-normal plasma temperature distribution model ($lnvmkl$) used in \citet{Ge2015}, for which we adopt the best-fit values for M32, i.e., peak temperature of $1.3$ keV, dispersion (in log-normal space) $\sigma _{\rm T}$ of $3.0$, fixed abundance at solar,
and a fixed normalization equivalent to a 0.5--2 keV emissivity of $6.0\times10^{27}\rm~erg~s^{-1}~M_\odot^{-1}$ \citep{Ge2015}.
We check the systematic uncertainty of this component by artificially increasing/decreasing the stellar emissivity by 30\%, a value estimated from \citet{Ge2015}. It is found that the related systematic uncertainty is small and has little effect in the best-fit hot gas temperature (see below) and the total flux. 
For the unresolved LMXBs, we adopt a power-law model with a photon-index fixed at 1.7 and set the normalization as a free parameter.
Finally, an {\it apec} model is adopted to account for the truly diffuse hot gas. The abundance parameter is fixed at solar, and all spectral components are subject to the Galactic foreground absorption column of $\rm N_H = 2\times10^{20}~cm^{-2}$.
For each galaxy, the spectra from individual observations are jointly fitted. We have adaptively binned the individual spectra to achieve a S/N greater than 3 per bin. 
VCC\,1630 has four available observations, but we neglect ObsID 8050, which provides only $\sim$15 net counts and only one effective bin.
We also neglect ObsID 8042 for VCC\,1030 for the same reason and fit the spectrum from the remaining single observation (ObsID 2883).

The composite model provides a reasonably good fit to all the spectra, as shown in Figure~\ref{fig:spectra}. 
In all five galaxies, a significant contribution arises from the {\it apec} component, indicating that truly diffuse hot gas exist in these galaxies.
The best-fit temperature of the hot gas is found to be $0.8^{+0.2}_{-0.6}$, $0.4^{+0.1}_{-0.1}$, $0.3^{+0.1}_{-0.1}$, and $0.3^{+0.1}_{-0.1}$ keV (90\% errors),
for VCC\,1030, VCC\,1154, VCC\,1231 and VCC\,1630, respectively. 
Empirically, these values are reasonable for the hot gas corona around intermediate-mass ETGs \citep[e.g.,][]{Boroson2011,Kim2015,Su2015}.
The case of VCC\,944 deserves some remarks. In the default model, the best-fit temperature of the hot gas is $0.1^{+0.2}_{-0.1}$ keV, whereas the power-law component dominates over almost the entire spectral range. However, the normalization of the power-law appears too high compared to the expected contribution from unresolved LMXBs, and a small contribution from the hot gas is also at odds with the very extended morphology of this galaxy (Figure~\ref{fig:fov_detected}). 
Therefore we also try an alternative model in which the normalization of the power-law component is fixed at the expected level of unresolved LMXBs, based on the stellar mass and source detection limit of VCC\,944. 
This second model results in an acceptable fit, in which the gas component dominates the spectrum, consistent with the observed morphology. However, the best-fit gas temperature is found to be $3.5^{+3.4}_{-1.3}$ keV, which is significantly higher than the typical temperature of the ICM in Virgo ($\sim$2 keV, \citealp{Urban2011}). If this relatively high temperature were real, it would imply a very high orbital velocity of VCC\,944.
A better understanding of the case of VCC\,944 would require deeper X-ray observations.
For VCC\,1030 and VCC\,1231, the power-law component has a negligible contribution compared to the other components. 
The spectral fit results of the five Virgo ETGs are listed in Table \ref{tab:spec}.

None of the seven field ETGs with detected diffuse X-ray emission have sufficient net counts for a reliable spectral analysis.



\section{Discussion} \label{sec:discussion}

Based on the above results, we now investigate the diffuse X-ray emission from our sample ETGs in the context of host galaxy mass and environment. A $L_{\rm X}-M_*$ diagram is instrumental for this purpose, which has been extensively examined in the literature.   
Our sample ETGs now allow us to extend such a diagram to the low-mass regime. 
To this end, we first convert the 0.5--2 keV photon fluxes (or 3$\sigma$ upper limits) into an unabsorbed luminosity ($L_{0.5-2}$).
For the five galaxies with good quality spectra (Section~\ref{subsec:spectra}),
we derive the 0.5--2 keV luminosity based on the best-fit spectral model, including both the stellar and gas components. 
For the other galaxies, since their X-ray emission is most likely dominated by unresolved stellar populations, we convert the 0.5--2 keV photon flux into intrinsic X-ray luminosity using a scaling factor of 
$2.0\times10^{41}{D}^2{\rm~erg~s^{-1}/(\rm photon~s^{-1}~cm^{-2})}$, where $D$ is the distance of a given galaxy in units of Mpc,
assuming an incident power-law spectrum with a photon-index of 1.7 (Section \ref{subsec:X-ray data}).
The derived 0.5--2 keV luminosity or $3\sigma$ upper limit of the Virgo ETGs are listed in Table~\ref{tab:info}.
The mean 0.5--2 keV luminosity or upper limit of each stacked galaxy bin is derived in the same way and listed in Table~\ref{tab:subset}.
For the stacked field ETGs, the converted luminosity assumes the median distance of a given mass bin.  
We have also compared a conversion factor assuming an optically-thin thermal model with a temperature of 0.3 keV and found that the resultant 0.5--2 keV luminosities are consistent to within $\sim 20\%$.

\begin{figure*}\centering
\includegraphics[width=0.7\textwidth,angle=90]{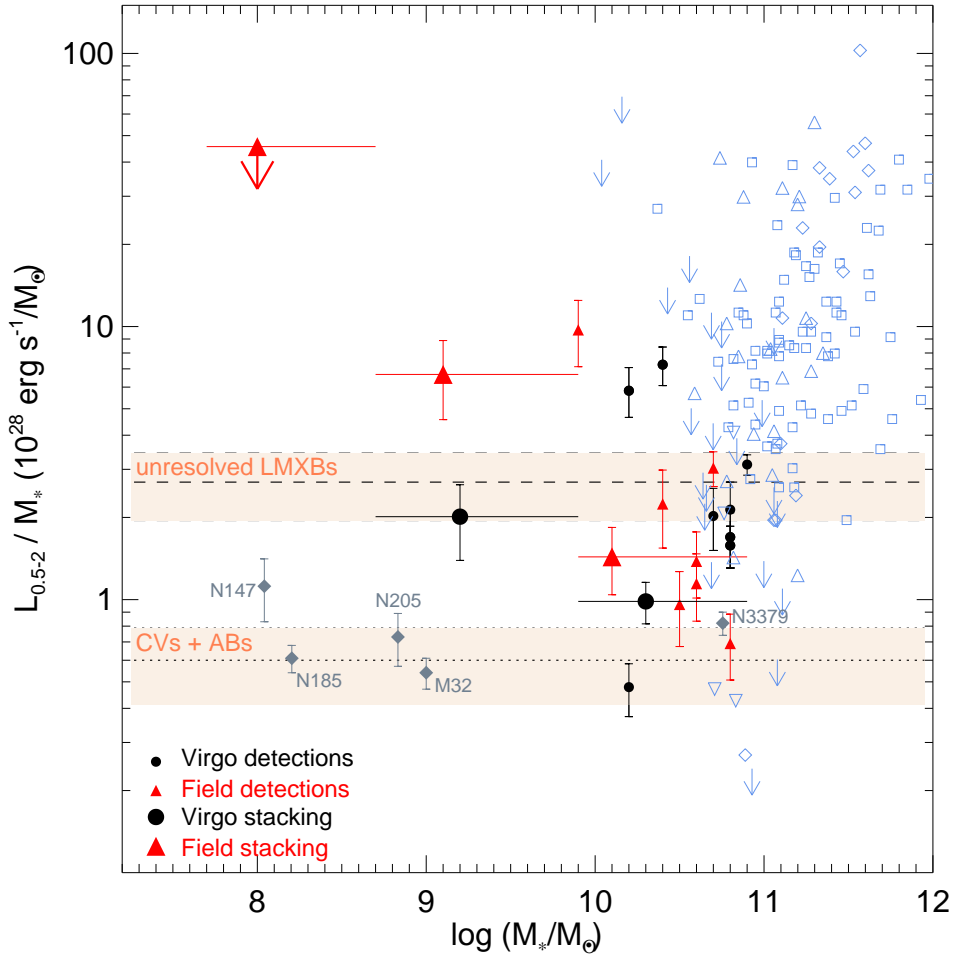}
\caption{
Specific 0.5--2 keV luminosity versus stellar mass of individually detected galaxies and the stacked mass bins. 
The small and large black circles in black are for Virgo detections and stacked bins, respectively.
The small and large red triangles are for field detections and stacked bins, respectively.  
The blue and gray symbols are literature measurements collected from \citet{Sun2007}, \citet{Mulchaey2010}, \citet{Jeltema2008}, \citet{Bogdan2012a}, \citet{Revnivtsev2008} and \citet{Ge2015}. 
The horizontal dashed line and dotted line indicate the expected contribution from unresolved LMXBs and the sum of CVs and ABs, respectively, with the uncertainties represented by the linen colored, horizontal belts. 
Downward pointing arrows represent upper limits.
}
\label{fig:lxmass}
\end{figure*}

Figure~\ref{fig:lxmass} shows the specific X-ray luminosity (i.e., $L_{0.5-2}/M_*$) versus stellar mass for our sample ETGs. Both   individually-detected galaxies and the stacked galaxy bins are plotted, with the Virgo ETGs in black and the AMUSE-Field ETGs in red.
The VLM bin of the field ETGs is plotted in 3$\sigma$ upper limits, due to its low S/N. 
For comparison, we also plot (in blue symbols) several samples of typically massive ETGs in the literature, which include  
ETGs in nearby galaxy clusters from \citet{Sun2007},
ETGs in nearby galaxy groups from \citet{Jeltema2008}, massive field ETGs from \citet{Mulchaey2010}.
When the $K$-band luminosity instead of stellar mass was given in the original study, we have converted the $K$-band luminosity into a stellar mass according to \citet{Bell2003} by assuming a typical $g-r$ of 0.7 for ETGs, which yields a mass-to-light ratio of $log_{10}(M_*/L)=0.07$.
We note that the stellar contribution has been subtracted from the 0.5--2 keV luminosity of these massive ETGs, i.e., the plotted values or upper limits are for diffuse hot gas only.
It can be seen that with our Virgo and field samples, we have significantly increased the number of individual X-ray measurements of ETGs with ${\rm log}M_* \approx 10.0-10.5$ and are able to probe the regime of ${\rm log}M_* \lesssim 9$.
To our knowledge, so far only four dwarf elliptical galaxies with such a low stellar mass, which include M32 \citep{Revnivtsev2007}, NGC\,147, NGC\,185 and NGC\,205 \citep{Ge2015}, have been closely examined with X-ray observations. 
We also plot these four dwarf ellipticals in Figure~\ref{fig:lxmass} (grey symbols), noting that their X-ray luminosities are dominated by CVs and ABs. The mean 0.5--2 keV emissivity of these four dwarf ellipticals, $0.6(\pm 0.2)\times10^{28}{\rm~erg~s^{-1}~M_\odot^{-1}}$, is shown by the horizontal dotted line and, as expected, defines a floor for our X-ray measurements.

In addition, we use a horizontal dashed line to indicate the potential contribution from unresolved LMXBs, which has a 0.5--2 keV emissivity of $2.7(\pm 0.7)\times10^{28}{\rm~erg~s^{-1}~M_\odot^{-1}}$. 
This value is obtained by adopting the LMXB luminosity function of \citet{Zhang2012} and
a power-law spectrum with a photon-index of 1.7 to convert the 0.5--8 keV luminosity used in that work into the 0.5--2 keV band. 
We have subtracted the contribution from LMXBs with a 0.5--8 keV luminosity above $\sim2\times 10^{38}{\rm~erg~s^{-1}}$, given the source detection limit for most of our sample ETGs.
For the few ETGs with deeper exposures, the contribution from unresolved LMXBs is expected to be much lower.


In Figure~\ref{fig:lxmass}, the massive (${\rm log}M_* \gtrsim 11$) ETGs show the familiar trend of an increasing X-ray luminosity with increasing stellar mass, albeit with large scatter. Except for a few non-detections, these massive ETGs must host substantial hot gas coronae. 
Our eight Virgo ETGs with detected diffuse X-ray emission appear to follow this general trend according to their stellar masses, although two galaxies, VCC\,944 and VCC\,1630, show a systematically higher emissivity than the other six galaxies.  
However, most intermediate-mass Virgo ETGs show no detectable diffuse X-ray emission. Their average X-ray emissivity, represented by the stacked IM signal, is consistent with being completely dominated by stellar populations, which is further supported by the radial intensity profile of the Virgo IM bin (Figure~\ref{fig:radial_stack}). 
The same is true for the low-mass Virgo ETGs. 
The stacked signals thus imply an upper limit of $10^{37-38}\rm~erg~s^{-1}$ for the X-ray luminosity of truly diffuse hot gas in these galaxies, which is at least a factor of 10 lower than that of the massive ETGs.
The apparent paucity of diffuse hot gas might be taken as a hint that most low- and intermediate-mass ETGs in Virgo have been stripped off of their hot gas content by ram pressure. 

However, our comparison sample of field ETGs also shows little evidence of diffuse X-ray emission. 
Indeed, all but 7 intermediate-mass field ETGs are also non-detections, and their stacked signal is consistent
with a pure stellar contribution and is comparable to that of the Virgo ETGs with similar stellar masses. 
Even among the 7 galaxies with detected diffuse X-ray emission, only two (NGC\,3193 and NGC\,3928) may harbor truly diffuse hot gas (Section~\ref{subsec:field}).
All field ETGs with ${\rm log}M_* \lesssim 10$ are also non-detections.
At face value, the stacked LM bin does show a mean specific X-ray luminosity ($6.7[\pm2.2] \times10^{28}{\rm~erg~s^{-1}~M_\odot^{-1}}$) marginally above the expected stellar contribution (LMXBs+CVs+ABs). However, the stacked signal of the field LM bin is weak and spatially concentrated (Figure~\ref{fig:stack_field}), and we find it premature to conclude in favor of the presence of truly diffuse hot gas in these galaxies. 
Therefore, we find no clear evidence that the Virgo ETGs have a systematically lower amount of diffuse hot gas compared to the field ETGs of similar stellar masses, as one would have naively expected if ram pressure stripping is effectively removing the hot gas in the former.
The deficiency of diffuse hot gas in the low- and intermediate-mass ETGs, both in Virgo and the field, may instead be due to their relatively shallow gravitational potentials, 
which are unable to hold much hot gas. Galactic winds launched by an AGN and/or Type Ia supernovae can expel the hot gas  from these galaxies, or at least from their inner regions probed by the current X-ray observations.
Thus, it is possible that in a large fraction of Virgo ETGs, ram pressure stripping is not required to keep the hot gas content at a low level.

Nevertheless, ram pressure stripping must have played a role in regulating the hot gas around the Virgo ETGs. 
M89 \citep{Machacek2006}, M86 \citep{Randall2008}, M49 \citep{Kraft2011}, M60 \citep{Wood2017} and NGC\,4342 \citep{Bogdan2012b} are known examples of Virgo ETGs with on-going ram pressure stripping.
Notably, with the exception of NGC\,4342, these are among the most massive galaxies in Virgo, thus their gravity is likely strong enough to compete with ram pressure and galactic wind in retaining a hot gas corona for a sufficiently long time. 
Among the 79 galaxies in our sample, 
VCC 944 (= NGC\,4417) and VCC 1154 (= NGC\,4459) stand out as good candidates of on-going ram pressure stripping (Figure~\ref{fig:fov_detected}), thanks to the relatively deep ACIS exposures of both galaxies.
Located at a projected radius of $\sim 2.9^{\circ}$ ($\sim$830 kpc, or about 80\% of the virial radius of Virgo) from M87 to the southwest,
NGC\,4417 is an S0 galaxy with ${\rm log}M_*=10.4$. 
A broad tail-like feature is present to the northwest side of the highly-inclined galactic disk, with an extent of at least $100\arcsec$ ($\sim$8 kpc), whereas there is no detectable diffuse emission on the southeast side. 
This extended and strongly lopsided morphology can only be understood as the presence of truly diffuse hot gas. 
The broad tail appears to emanate from the galactic disk, reminiscent of HI tails in ram pressure stripping spiral galaxies, but here the host galaxy is an S0 that should not contain much cold gas. 
In fact, our inspection of archival HST/CFHT/2MASS images finds no obvious optical/near-infrared counterpart for the X-ray tails, nor did we find clear evidence of cold gas or dust within the disk.

A similar morphology is witnessed in NGC\,4459, which is also an S0 galaxy and one of the largest galaxies (${\rm log}M_*=10.9$) in our sample, located at a projected radius of $\sim 1.7^{\circ}$ ($\sim$480 kpc, or about 45\% of the virial radius of Virgo) from M87 to the northwest.
The diffuse emission extends towards the south of the galaxy, where a tail-like feature slightly swings to the southeast, reaching to a projected distance of $\sim100\arcsec$. 
\citet{Juranova2020} also presented evidence for this tail-like feature with XMM-Newton observations, albeit at a lower resolution.
An even more puzzling diffuse feature is seen $\sim2\arcmin$ away from NGC\,4459 to the northwest, which has the shape of a wide filament, with a size of $\sim2\arcmin \times 1\farcm5$  (Figure~\ref{fig:fov_detected}).
We find no obvious optical counterpart at the position of this feature.
The spectrum of this feature can be well described by an optically-thin plasma model, but requiring a none-zero redshift of $z = 0.57 \pm 0.20$ and a gas temperature of $5.3\pm1.5$ keV (90\% errors).
These values, together with the substantial extent, suggest that this feature might be a background galaxy cluster.
Deeper X-ray/optical observations are needed to determine the nature of this diffuse feature.

It has been suggested that ram pressure may help to halt galactic outflows and temporarily produce pressure-confined coronae with enhanced gas density and diffuse X-ray luminosity \citep[e.g.,][]{Brown2000}. 
Such an effect has been studied with hydrodynamic simulations by \citet{Lu2011}, who took into account stellar mass loss and energy input from Type Ia SNe in a simulated galaxy with ${\rm log}M_* = 11$ moving through the ICM of representative conditions. 
They found that a sizable hot gas corona can be maintained under the combined effects of ram pressure and stellar feedback.
The X-ray luminosities of their simulated hot gas coronae range between a few $10^{37}{\rm~erg~s^{-1}}$ to a few $10^{39}{\rm~erg~s^{-1}}$.
For smaller galaxies similar to our sample, the X-ray luminosity of a pressure-confined corona would be even lower. 
Unfortunately, the sensitivity of our current observations is not sufficient to probe such ram pressure-confined coronae except for the most luminous ones.
We point out that one such case might have been found with NGC\,4342, an  intermediate-mass (${\rm log}M_* \approx 10.3$) S0 galaxy lying at the southwest outskirt of Virgo. The X-ray morphology of NGC\,4342 exhibits a sharp cold front and a trailing tail, both characteristic of on-going ram pressure stripping \citep{Bogdan2012b}. The measured diffuse X-ray luminosity of NGC\,4342, $\sim3\times10^{39}{\rm~erg~s^{-1}}$, is exceptionally high for its current stellar mass, but may be understood in the sense of a semi-confined corona studied by \citet{Lu2011}.
We note that the morphology of NGC\,4459 may also be interpreted as having a semi-confined corona (to the north), although a sharp cold front is not unambiguously seen due to the limited S/N.

Lastly, we note that the stacked Virgo IM bin has a specific X-ray luminosity of $\sim1.0\times10^{28}{\rm~erg~s^{-1}~M_\odot^{-1}}$, which is significantly lower than the expected contribution from unresolved LMXBs and also lower than that of the stacked field IM bin (Figure~\ref{fig:lxmass}).
This deviation deserves some remarks. 
A simple explanation is that in the Virgo intermediate-mass ETGs, the specific number of low-luminosity LMXBs is lower than the standard luminosity function of \citet{Zhang2012} predicts.   
LMXBs are generally thought to have two main origins: LMXBs formed in the field following binary evolution in isolation, and LMXBs dynamically formed in high stellar density environments, especially in globular clusters \citep[GCs;][]{Clark1975,Fabbiano2006}. 
\citet{Zhang2012} found that the specific number of LMXBs in ETGs depends on both galaxy age and the specific frequency of GCs, which can vary by a factor of up to $\sim$2 from the mean value. 
An example of low specific number of LMXBs is found with NGC\,3379, which is highlighted in Figure~\ref{fig:lxmass}. 
This relatively massive elliptical galaxy has one of the lowest specific X-ray luminosities for its mass, which was suggested to be dominated by unresolved stars \citep{Revnivtsev2008}. 
\citet{Zhang2012} found that NGC\,3379 has a specific number of LXMBs about two times lower than the mean, which may be related to its low GC specific frequency, defined as $S_N \equiv N_{\rm GC}10^{0.4(M^{\rm T}_{\rm V}+15)}$, where $M^{\rm T}_{\rm V}$ is the $V$-band absolute magnitude and $N_{\rm GC}$ is the total number of GCs. 
We cross-correlate the \citet{Harris2013} catalog of local galaxies with published GC properties to derive the GC specific frequency for our sample ETGs. 
Indeed, 59 Virgo intermediate-mass ETGs listed in \citet{Harris2013} have $S_N < 3$, lower than the mean value of the \citet{Zhang2012} sample. 
Hence this may be the cause of the low specific X-ray luminosity of the Virgo IM bin.

\section{Summary} \label{sec:summary}
We have presented a systematic study of diffuse hot gas around 79 low-to-intermediate stellar mass ($M_* \approx 10^{9-11}\rm~M_\odot$) ETGs in the Virgo cluster as well as a comparison sample of 57 ETGs in the field, using archival {\it Chandra} observations. The main results are as follows:
\begin{itemize}
\item We detect 0.5--2 keV diffuse X-ray emission in only eight Virgo ETGs (all with stellar mass $M_* > 10^{10}\rm~M_{\odot}$), and in only five cases can a substantial fraction of the detected signals  be unambiguously attributed to truly diffuse hot gas, based on their spatial distributions and spectral properties. The remaining 71 low-to-intermediate mass ETGs do not have significant diffuse X-ray emission (S/N $<$ 3).

\item For the individually non-detected ETGs, we constrain their average X-ray emission with a stacking analysis. The specific X-ray luminosity of $L_{\rm X}/M_* \sim 10^{28}{\rm~erg~s^{-1}~M_{\odot}^{-1}}$ is consistent with the expected contribution from unresolved stellar populations (LMXBs, CVs and ABs). 
The apparent paucity of truly diffuse hot gas in these low- and intermediate-mass ETGs may be understood as efficient ram-pressure stripping by the hot intra-cluster medium. 

\item We find strong morphological evidence for on-going ram pressure stripping in two galaxies (NGC\,4417 and NGC\,4459) in the Virgo cluster.

\item As for the field ETGs with a similar detection sensitivity, both the individually-detected and stacked signals of diffuse X-ray emission are comparable to that of the Virgo ETGs of similar stellar masses and are consistent with an unresolved stellar origin. Since field galaxies are not expected to suffer from strong ram pressure stripping, the paucity of hot gas in the field ETGs may be regarded as evidence that a galactic wind driven by nuclear activities and/or Type Ia supernovae is the primary mechanism of evacuating hot gas from the inner region of low- and intermediate-mass ETGs. Galactic winds may also play a partial role in reducing the hot gas content of the Virgo ETGs.

\end{itemize}

The on-going eROSITA all-sky survey may help significantly advance our understanding of the roles of ram pressure stripping and galactic winds in regulating the hot gas content of ETGs in the local universe.

\acknowledgments
This research has made use of data and software provided by the {\it Chandra X-ray Observatory}.
M.H. is grateful to the hospitality of the Smithsonian Astrophysical Observatory where part of this work was conducted, and would like to thank Ralph Kraft and {\'A}kos Bogd{\'a}n for helpful discussions.
M.H. and Z.L. acknowledge support by the National Key Research and Development Program of China (grant 2017YFA0402703).

\clearpage
\startlongtable
\begin{deluxetable}{ccccccccc}
\tabletypesize{\footnotesize}
\tablecaption{Basic information of the Virgo early-type galaxies} \label{tab:info}
\tablewidth{0pt}
\tablehead{
\colhead{Galaxy name} & \colhead{Other name} & \colhead{ObsID} & \colhead{R.A.} & \colhead{Dec.} &
\colhead{Exp.} & \colhead{${R_{\rm e}}$} & \colhead{log $M_*$} & \colhead{$L_{\rm X, {diff}}$}
}
\colnumbers
\startdata
    VCC\,9  &    IC\,3019  & 8072  & 182.34275  &  13.99243  &   5.3  &  29.62  &  9.7  &  $<$ 14.0 \\
   VCC\,21  &    IC\,3025  & 8089  & 182.59621  &  10.18854  &   5.1  &   9.76  &  9.0  &  $<$  3.1 \\
   VCC\,33  &    IC\,3032  & 8086  & 182.78234  &  14.27481  &   4.5  &   9.10  &  8.9  &  $<$  4.7 \\
  VCC\,140  &    IC\,3065  & 8076  & 183.80234  &  14.43288  &   5.1  &   9.24  &  9.4  &  $<$  3.2 \\
  VCC\,200  &         ...  & 8087  & 184.14044  &  13.03157  &   5.1  &  12.80  &  9.2  &  $<$  4.5 \\
  VCC\,230  &    IC\,3101  & 8100  & 184.33188  &  11.94347  &   5.1  &   9.71  &  8.9  &  $<$  3.7 \\
  VCC\,355  &   NGC\,4262  & 8049  & 184.87740  &  14.87766  &   3.1  &   9.78  & 10.3  &  $<$  7.6 \\
  VCC\,369  &   NGC\,4267  & 8039  & 184.93848  &  12.79828  &   5.1  &   7.93  & 10.4  &  $<$  6.6 \\
  VCC\,437  &  UGC\,7399A  & 8085  & 185.20341  &  17.48707  &   5.1  &  25.02  &  9.6  &  $<$ 12.8 \\
  VCC\,538  &  NGC\,4309A  & 7122 {\it 8105}  & 185.56143  &   7.16715  &   8.6  &   4.82  &  8.9  &  $<$  2.6 \\
  VCC\,543  &   UGC\,7436  & 8080  & 185.58137  &  14.76078  &   5.3  &  17.14  &  9.4  &  $<$  8.0 \\
  VCC\,571  &         ...  & 8088  & 185.67149  &   7.95035  &   5.1  &  13.26  &  9.4  &  $<$  4.2 \\
  VCC\,575  &   NGC\,4318  & 8073  & 185.68039  &   8.19828  &   5.1  &  17.85  & 10.8  &  $<$  6.2 \\
  VCC\,654  &   NGC\,4340  & 8045  & 185.89703  &  16.72236  &   5.1  &  21.63  & 10.4  &  $<$ 13.3 \\
  VCC\,685  &   NGC\,4350  & 4015  & 185.99111  &  16.69341  &   3.3  &  11.91  & 10.6  &  $<$  9.0 \\
  VCC\,698  &   NGC\,4352  & 8068  & 186.02094  &  11.21807  &   5.1  &  16.01  & 10.0  &  $<$  4.5 \\
  VCC\,751  &    IC\,3292  & 8103  & 186.20150  &  18.19512  &   5.1  &   9.95  &  9.4  &  $<$  4.6 \\
  VCC\,759  &   NGC\,4371  & 8040  & 186.23096  &  11.70421  &   4.9  &  26.18  & 10.8  &  $ 13.5 \pm 3.5$  \\
  VCC\,778* &   NGC\,4377  & 8055 13295 15783 16476  & 186.30140  &  14.76216  & 132.4  &   5.90  & 10.2  &  $  0.8 \pm 0.2$  \\
            &              & 16477 17558 17559   &  &  &  &  &  &  \\
  VCC\,784  &   NGC\,4379  & 8053  & 186.31142  &  15.60747  &   5.1  &  14.24  & 10.3  &  $<$  4.7 \\
  VCC\,856  &    IC\,3328  & 8128  & 186.49136  &  10.05377  &   5.1  &  14.72  &  9.5  &  $<$  5.7 \\
  VCC\,944* &   NGC\,4417  & 8125 14902  & 186.71088  &   9.58426  &  34.8  &  12.57  & 10.4  &  $ 18.1 \pm 2.4$  \\
 VCC\,1025  &   NGC\,4434  & 8060  & 186.90285  &   8.15434  &   5.1  &  11.13  & 10.4  &  $<$ 10.0 \\
 VCC\,1030* &   NGC\,4435  & 2883 8042  & 186.91870  &  13.07894  &  26.7  &  16.42  & 10.8  &  $ 10.4 \pm 2.2$  \\
 VCC\,1049  &   UGC\,7580  & 8075  & 186.97847  &   8.09040  &   5.5  &   9.94  &  9.0  &  $<$  5.1 \\
 VCC\,1062  &   NGC\,4442  & 8037  & 187.01618  &   9.80371  &   5.3  &  16.92  & 10.7  &  $ 10.1 \pm 2.7$  \\
 VCC\,1075  &    IC\,3383  & 8096  & 187.05134  &  10.29766  &   5.1  &  14.57  &  9.2  &  $<$  3.8 \\
 VCC\,1087  &    IC\,3381  & 8078  & 187.06202  &  11.78983  &   5.1  &  18.87  &  9.6  &  $<$  8.9 \\
 VCC\,1125  &   NGC\,4452  & 8064  & 187.18045  &  11.75503  &   5.1  &  13.26  &  9.9  &  $<$  6.1 \\
 VCC\,1146  &   NGC\,4458  & {\it 8059} 14905  & 187.23986  &  13.24189  &  34.1  &  18.60  & 10.0  &  $<$ 10.9 \\
 VCC\,1154* &   NGC\,4459  & 2927 11784  & 187.25004  &  13.97837  &  39.6  &  28.15  & 10.9  &  $ 24.8 \pm 2.0$  \\
 VCC\,1178  &   NGC\,4464  & 8127  & 187.33872  &   8.15662  &   5.1  &   6.58  &  9.9  &  $<$  6.2 \\
 VCC\,1231${}^\dag$ &   NGC\,4473  & 4688  & 187.45363  &  13.42936  &  29.6  &  16.89  & 10.8  &  $  9.8 \pm 1.6$  \\
 VCC\,1242  &   NGC\,4474  & 8052  & 187.47311  &  14.06859  &   5.1  &  16.88  & 10.3  &  $<$ 11.1 \\
 VCC\,1261  &   NGC\,4482  & 8067  & 187.54303  &  10.77948  &   5.1  &  20.95  &  9.8  &  $<$  5.0 \\
 VCC\,1283  &   NGC\,4479  & 8066  & 187.57655  &  13.57764  &   5.1  &  19.28  & 10.1  &  $<$  7.4 \\
 VCC\,1303  &   NGC\,4483  & 8061  & 187.66936  &   9.01568  &   5.1  &  15.18  & 10.1  &  $<$  5.3 \\
 VCC\,1321  &   NGC\,4489  & 8126  & 187.71771  &  16.75885  &   5.1  &  37.07  & 10.1  &  $<$ 13.2 \\
 VCC\,1355  &    IC\,3442  & 8077  & 187.83414  &  14.11520  &   4.8  &  24.55  &  9.4  &  $<$  8.5 \\
 VCC\,1422  &    IC\,3468  & 8069  & 188.05922  &  10.25146  &   5.1  &  19.29  &  9.6  &  $<$  6.7 \\
 VCC\,1431  &    IC\,3470  & 8081  & 188.09745  &  11.26297  &   5.1  &   9.86  &  9.5  &  $<$  5.5 \\
 VCC\,1440  &     IC\,798  & 8099  & 188.13919  &  15.41539  &   5.5  &   7.75  &  9.2  &  $<$  3.5 \\
 VCC\,1475  &   NGC\,4515  & 8065  & 188.27073  &  16.26553  &   5.1  &   9.49  &  9.9  &  $<$  4.7 \\
 VCC\,1488  &    IC\,3487  & 8090  & 188.30598  &   9.39735  &   5.1  &   9.31  &  9.0  &  $<$  4.0 \\
 VCC\,1489  &    IC\,3490  & 8113  & 188.30793  &  10.92854  &   5.1  &  10.40  &  8.7  &  $<$  5.4 \\
 VCC\,1499  &    IC\,3492  & 8093  & 188.33241  &  12.85343  &   5.1  &   5.50  &  8.8  &  $<$  4.7 \\
 VCC\,1512  &         ...  & 8112  & 188.39440  &  11.26196  &   5.1  &  12.79  &  9.2  &  $<$  3.6 \\
 VCC\,1528  &    IC\,3501  & 8082  & 188.46508  &  13.32244  &   4.8  &   9.90  &  9.3  &  $<$  5.0 \\
 VCC\,1537  &   NGC\,4528  & 8054  & 188.52531  &  11.32126  &   5.1  &   8.01  & 10.1  &  $<$  8.4 \\
 VCC\,1539  &    IC\,3506  & 8109  & 188.52808  &  12.74160  &   5.1  &  26.30  &  8.9  &  $<$ 21.2 \\
 VCC\,1545  &    IC\,3509  & 8094  & 188.54806  &  12.04896  &   5.1  &  11.24  &  9.2  &  $<$  4.8 \\
 VCC\,1619  &   NGC\,4550  & {\it 8050} 8058 16032 16033 & 188.87743  &  12.22082  &  50.1  &  10.43  & 10.2  &  $<$  5.2 \\
 VCC\,1627  &         ...  & 8058 {\it 8098} 16033  & 188.90521  &  12.38203  &  31.6  &   3.73  &  9.1  &  $<$  1.6 \\
 VCC\,1630* &   NGC\,4551  & 8050 8058 16032 16033  & 188.90814  &  12.26398  &  50.1  &  13.29  & 10.2  &  $  9.4 \pm 1.7$  \\
 VCC\,1661  &         ...  & 8044  & 189.10327  &  10.38466  &   4.9  &  58.07  &  9.0  &  $<$ 28.9 \\
 VCC\,1692  &   NGC\,4570  & 8041  & 189.22250  &   7.24664  &   5.1  &   9.50  & 10.6  &  $<$  9.2 \\
 VCC\,1695  &    IC\,3586  & 8083  & 189.22853  &  12.52007  &   5.1  &  20.21  &  9.5  &  $<$  9.7 \\
 VCC\,1720  &   NGC\,4578  & 8048  & 189.37734  &   9.55507  &   5.1  &  32.52  & 10.4  &  $<$ 19.2 \\
 VCC\,1743  &         ...  & 8108  & 189.52830  &  10.08235  &   5.1  &  10.41  &  8.9  &  $<$  3.7 \\
 VCC\,1779  &    IC\,3612  & 8091  & 189.76960  &  14.73113  &   5.1  &  10.77  &  9.0  &  $<$  5.0 \\
 VCC\,1826  &    IC\,3633  & 8111  & 190.04688  &   9.89613  &   5.1  &   7.06  &  8.8  &  $<$  6.1 \\
 VCC\,1828  &    IC\,3635  & 8104  & 190.05573  &  12.87477  &   5.1  &  14.55  &  9.1  &  $<$  4.6 \\
 VCC\,1833  &         ...  & 8084  & 190.08199  &  15.93528  &   5.2  &   7.38  &  9.3  &  $<$  3.9 \\
 VCC\,1857  &    IC\,3647  & 8130  & 190.22131  &  10.47536  &   5.1  &  20.82  &  9.0  &  $<$ 11.3 \\
 VCC\,1861  &    IC\,3652  & 8079  & 190.24400  &  11.18449  &   5.1  &  15.14  &  9.5  &  $<$  6.5 \\
 VCC\,1871  &    IC\,3653  & 8071  & 190.31556  &  11.38725  &   5.1  &   6.82  &  9.5  &  $<$  3.2 \\
 VCC\,1883  &   NGC\,4612  & 8051  & 190.38646  &   7.31488  &   5.1  &  24.99  & 10.4  &  $<$ 10.5 \\
 VCC\,1886  &    IC\,3663  & 8106  & 190.41422  &  12.24739  &   5.1  &  12.61  &  8.8  &  $<$  4.3 \\
 VCC\,1895  &   UGC\,7854  & 8092  & 190.46658  &   9.40289  &   5.1  &  10.07  &  9.0  &  $<$  4.0 \\
 VCC\,1910  &     IC\,809  & 2068 {\it 8074}  & 190.53610  &  11.75429  &  30.1  &  12.52  &  9.5  &  $<$  4.2 \\
 VCC\,1913  &   NGC\,4623  & 8062  & 190.54454  &   7.67694  &   5.3  &  14.67  & 10.1  &  $<$  4.0 \\
 VCC\,1938  &   NGC\,4638  & 8046  & 190.69760  &  11.44251  &   5.3  &  14.79  & 10.5  &  $<$ 11.1 \\
 VCC\,1948  &    IC\,3693  & 8097  & 190.74175  &  10.68181  &   5.1  &  10.94  &  8.8  &  $<$  4.5 \\
 VCC\,1993  &         ...  & 8102  & 191.05011  &  12.94184  &   4.8  &   8.58  &  9.0  &  $<$  4.9 \\
 VCC\,2000  &   NGC\,4660  & 8043  & 191.13327  &  11.19053  &   5.1  &  10.45  & 10.4  &  $<$  8.6 \\
 VCC\,2019  &    IC\,3735  & 8129  & 191.33510  &  13.69266  &   3.5  &  15.25  &  9.4  &  $<$  8.5 \\
 VCC\,2048  &    IC\,3773  & 8070  & 191.81375  &  10.20359  &   5.3  &  11.90  &  9.6  &  $<$  4.6 \\
 VCC\,2050  &    IC\,3779  & 8101  & 191.83598  &  12.16643  &   5.1  &  10.80  &  9.0  &  $<$  4.9 \\
 VCC\,2092  &   NGC\,4754  & 8038  & 193.07290  &  11.31389  &   3.4  &  24.13  & 10.9  &  $<$ 22.1 \\
\enddata
\tablecomments{(1) Name of Virgo galaxies (* denotes galaxies with multiple observations which result in detection of diffuse X-ray emission, ${}^\dag$ denotes the only one galaxy VCC\,1231 with single longer observation);
(2) Other name of the target galaxies; (3) {\it Chandra} observation ID. For galaxies with multiple observations, all observation ID are listed, and the 5-ks observation used in stacking analysis is marked in italic; (4)-(5): Celestial coordinates of the galactic center (J2000); (6) {\it Chandra} effective exposure, in units of ks; (7) Effective radius of the target galaxies, in units of arc-second; (8) Logarithmic stellar mass of host galaxies, in unit of $M_\odot$, adopted from Gallo et al. (2010); (9) 0.5-2 keV luminosity of detected diffuse X-ray emission and $3\sigma$ upper limit of non-detections, in units of ${10^{38}\rm~erg~s^{-1}}$.
}
\end{deluxetable}

\clearpage
\begin{deluxetable}{cccccccc}
\tabletypesize{\footnotesize}
\tablecaption{Stacking results of various galaxy subsets}
\label{tab:subset}
\tablehead{
\colhead{Subset} & \colhead{\# of ETGs} & \colhead{Mass Range} & \colhead{Median $M_*$} & \colhead{S/N} & 
\colhead{Photon Flux} &
\colhead{$L_{\rm X}$}
& \colhead{$L_{\rm X}/M_*$} 
}
\colnumbers
\startdata
Intermediate-mass Virgo ETGs& 25 & $9.9 \leq M_* \leq 10.9$&  10.3& 5.7& $3.6 \pm 0.6$ & $2.0 \pm 0.3$ & $1.0 \pm 0.2$  \\ 
Low-mass Virgo ETGs& 45 & $8.7 \leq M_* < 9.9$ & 9.2& 3.2& $0.6 \pm 0.2$ &  $0.3 \pm 0.1$ & $2.0 \pm 0.6$ \\ 
Intermediate-mass field ETGs& 7 & $9.9 < M_* < 10.9$&  10.1& 3.6& $2.3 \pm 0.6$ & $1.8 \pm 0.5$& $1.4 \pm 0.4$ \\ 
Low-mass field ETGs&  15 & $8.7 \leq M_* < 9.9$& 9.1 & 3.1& $0.9 \pm 0.3$ & $0.8 \pm 0.3$ & $6.7 \pm 2.2$ \\ 
Very low-mass field ETGs& 27 & $7.7 \leq M_* < 8.7$ & 8.0& 0.3& $<$ 0.4 & $<$ 0.5 & $<$ 45.6 \\ 
\enddata
\tablecomments{(1) Subsets of stacked galaxies; (2) Number of stacked galaxies; (3) Logarithmic stellar mass range of stacked galaxies; (4) Median stellar mass of stacked galaxies; (5) Signal-to-noise ratio of the stacked signal; (6) Net photon flux per stacked galaxy, in units of ${10^{-6}\rm~ph~s^{-1}~cm^{-2}}$; (7) Mean 0.5--2 keV luminosity of the stacked signal in units of ${10^{38}\rm~erg~s^{-1}}$, converted from column 6 by adopting the median distance of a given bin; (8) Specific 0.5--2 keV luminosity of the stacked signal, in units of ${10^{28}\rm~erg~s^{-1}~M_\odot^{-1}}$. The upper limits are of 3$\sigma$.}
\end{deluxetable}

\begin{deluxetable}{ccccccc}
\tabletypesize{\footnotesize}
\tablecaption{Spectral fit results of five Virgo ETGs}
\label{tab:spec}
\tablehead{
\colhead{Name} & \colhead{$kT_{\rm gas}$} & \colhead{$\chi^2/$d.o.f.} & \colhead{$\rm Flux_{(gas)}$} & \colhead{$\rm Flux_{(CV+AB)}$} & \colhead{$\rm Flux_{(LMXB)}$} & \colhead{Total flux}
}
\colnumbers
\startdata
VCC\,944$^*$  & $0.1^{+0.2}_{-0.1}$ & 21.9/16 & $1.0 \pm 0.7$ & $0.7 \pm 0.6$ & $4.0 \pm 0.6$ & $5.5 \pm 0.7$ \\
& $3.5^{+3.4}_{-1.3}$ & 25.8/17 & $3.4 \pm 0.6$ & $0.8 \pm 0.6$ & $0.8 \pm 0.6$ & $4.7 \pm 0.6$ \\
VCC\,1030 & $0.8^{+0.2}_{-0.6}$ & 4.4/4  & $1.5 \pm 0.6$ & $1.5 \pm 0.7$ & $0.3^{+0.6}_{-0.3}$ & $3.2 \pm 0.7$ \\
VCC\,1154 & $0.4^{+0.1}_{-0.1}$ & 31.0/31 & $5.0 \pm 0.5$ & $1.4 \pm 0.6$ & $1.1 \pm 0.6$ & $7.6 \pm 0.6$ \\
VCC\,1231 & $0.3^{+0.1}_{-0.1}$ & 10.4/7  & $1.5 \pm 0.4$ & $1.5 \pm 0.5$ & $0.05^{+0.51}_{-0.05}$ & $3.0 \pm 0.5$ \\
VCC\,1630 & $0.3^{+0.1}_{-0.1}$ & 4.7/7  & $2.1 \pm 0.5$ & $0.4 \pm 0.4$ & $0.4 \pm 0.3$ & $2.9 \pm 0.5$ \\
\enddata
\tablecomments{(1) Galaxy name; (2) Best-fit temperature of hot gas; (3) $\chi^2$ over degree-of-freedom; (4)-(6) 0.5--2 keV flux of different components derived from best-fit spectral model, in units of ${10^{-14}\rm~erg~s^{-1}~cm^{-2}}$; (7) Total 0.5--2 keV flux, including all three components, in units of ${10^{-14}\rm~erg~s^{-1}~cm^{-2}}$. Quoted errors are at 90\% confidence level. $^*$This galaxy is fitted with two models, in the first case the flux of the LMXB component is set free, while in the second case the flux of the LMXB component is fixed at the expected level.}
\end{deluxetable}


\bibliography{diffuserefs}{}
\bibliographystyle{aasjournal}


\end{document}